\newif\ifsubmode
\submodefalse

\newif\ifcmtmode
\cmtmodetrue

\documentclass{emulateapj}
\usepackage{natbib,amsmath,verbatim}

\usepackage{rotating}
\usepackage[usenames,dvipsnames]{color}

\usepackage{amsmath}
\usepackage{longtable}

\newcommand{\refbf}{}
\newcommand{\refbfr}{}
\newcommand{\refbfb}{}
\newcommand{\refbfg}{}
\newcommand{\refg}{}
\newcommand{\refc}{}

\newcommand{\refq}{}
\newcommand{\refr}{}
\newcommand{\refrr}{}
\newcommand{\refrrr}{}
\newcommand{\refrrrr}{}

\newcommand{\refrrrrr}{}
\newcommand{\refx}{}
\newcommand{\refxx}{}
\newcommand{\refxxx}{}
\newcommand{\refxxxx}{}

\usepackage{hyperref}

\shorttitle{The J1713+0747 24-Hour Global Campaign}
\shortauthors{Dolch \& Lam et al.}
\submitted{Astrophysical Journal, 794: 21, (Received 2014 July 5; accepted 2014 August 5; published 2014 October 10)}

\begin{document}

\title{A 24-Hour Global Campaign to Assess Precision Timing of the Millisecond Pulsar J1713+0747}
\author{ 
T.\,Dolch\altaffilmark{1,2}, 
M.\,T.\,Lam\altaffilmark{1},
J.\,Cordes\altaffilmark{1},
S.\,Chatterjee\altaffilmark{1},  
C.\,Bassa\altaffilmark{3,4},
B.\,Bhattacharyya\altaffilmark{4,5},
D.\,J.\,Champion\altaffilmark{6},
I.\,Cognard\altaffilmark{7},
K.\,Crowter\altaffilmark{8},
P.\,B.\,Demorest\altaffilmark{9},
J.\,W.\,T.\,Hessels\altaffilmark{3,10},
G.\,Janssen\altaffilmark{3,4},
F.\,A.\,Jenet\altaffilmark{11},
G.\,Jones\altaffilmark{12},
C.\,Jordan\altaffilmark{4},
R.\,Karuppusamy\altaffilmark{6},
M.\,Keith\altaffilmark{4},
V.\,Kondratiev\altaffilmark{3,13},
M.\,Kramer\altaffilmark{6,14},
P.\,Lazarus\altaffilmark{6},
T.\,J.\,W.\,Lazio\altaffilmark{15},
K.\,J.\,Lee\altaffilmark{16,6},
M.\,A.\,McLaughlin\altaffilmark{17},
J.\,Roy\altaffilmark{4,5},
R.\,M.\,Shannon\altaffilmark{18},
I.\,Stairs\altaffilmark{8},
K.\,Stovall\altaffilmark{19},
J.\,P.\,W.\,Verbiest\altaffilmark{20,6},
D.\,R.\,Madison\altaffilmark{1},
N.\,Palliyaguru\altaffilmark{17},
D.\,Perrodin\altaffilmark{21},
S.\,Ransom\altaffilmark{9},
B.\,Stappers\altaffilmark{4},
W.\,W.\,Zhu\altaffilmark{8},
S.\,Dai\altaffilmark{22,18},
G.\,Desvignes\altaffilmark{6},
L.\,Guillemot\altaffilmark{7},
K.\,Liu\altaffilmark{7},
A.\,Lyne\altaffilmark{4},
B.\,B.\,P.\,Perera\altaffilmark{4},
E.\,Petroff\altaffilmark{18,23,24},
J.\,M.\,Rankin\altaffilmark{25}, 
R.\,Smits\altaffilmark{3}}
\altaffiltext{1}{Astronomy Department, Cornell University, Ithaca, NY 14853, USA; tdolch@astro.cornell.edu}
\altaffiltext{2}{Department of Physics and Astronomy, Oberlin College, Oberlin, OH 44074, USA}
\altaffiltext{3}{ASTRON, the Netherlands Institute for Radio Astronomy, Postbus 2, 7990 AA, Dwingeloo, The Netherlands}
\altaffiltext{4}{Jodrell Bank Centre for Astrophysics, School of Physics and Astronomy, The University of Manchester, Manchester M13 9PL, UK}
\altaffiltext{5}{National Centre for Radio Astrophysics, Tata Institute of Fundamental Research, Pune 411007, India.}
\altaffiltext{6}{Max-Planck-Institut f\"ur Radioastronomie, Auf dem H\"ugel 69, D-53121 Bonn, Germany}
\altaffiltext{7}{Laboratoire de Physique et Chimie de l'Environnement et de l'Espace, LPC2E UMR 6115 CNRS, F-45071 Orl\'eans Cedex 02, and Station de radioastronomie de Nan\c{c}ay, Observatoire de Paris, CNRS/INSU, F-18330 Nan\c{c}ay, France}
\altaffiltext{8}{Department of Physics and Astronomy, University of British Columbia, 6224 Agricultural Road, Vancouver, BC V6T 1Z1, Canada}
\altaffiltext{9}{National Radio Astronomy Observatory, 520 Edgemont Road, Charlottesville, VA 22901, USA}
\altaffiltext{10}{Anton Pannekoek Institute for Astronomy, University of Amsterdam, Science Park 904, 1098 XH Amsterdam, The Netherlands}
\altaffiltext{11}{Center for Advanced Radio Astronomy, University of Texas, Rio Grande Valley, Brownsville, TX 78520, USA}
\altaffiltext{12}{Columbia Astrophysics Laboratory, Columbia University, NY 10027, USA}
\altaffiltext{13}{Astro Space Center of the Lebedev Physical Institute, Profsoyuznaya str. 84/32, Moscow 117997, Russia}
\altaffiltext{14}{University of Manchester, Jodrell Bank Observatory, Macclesfield, Cheshire, SK11 9DL, UK}
\altaffiltext{15}{Jet Propulsion Laboratory, California Institute of Technology, 4800 Oak Grove Drive, Pasadena, CA 91106, USA}
\label{firstpage}
\begin{abstract}

The radio millisecond pulsar J1713+0747 is regarded as one of the highest-precision clocks in the sky, and is regularly timed for the purpose of detecting gravitational waves. The International Pulsar Timing Array collaboration undertook a 24-hour global observation of PSR~J1713+0747 in an effort to better quantify sources of timing noise in this pulsar, particularly on intermediate (1 -- 24\,hr) timescales. We observed the pulsar continuously over 24\,hr with the Arecibo, Effelsberg, GMRT, Green Bank, LOFAR, Lovell, Nan\c cay, Parkes, and WSRT radio telescopes. The combined pulse times-of-arrival presented here provide an estimate of what sources of timing noise, excluding DM variations, would be present as compared to an idealized $\sqrt{N}$ improvement in timing precision, where $N$ is the number of pulses analyzed. In the case of this particular pulsar, we find that intrinsic pulse phase jitter dominates arrival time precision when the S/N of single pulses exceeds unity, as measured using the eight telescopes that observed at L-band/1.4\,GHz. We present first results of specific phenomena probed on the unusually long timescale (for a single continuous observing session) of tens of hours, in particular interstellar scintillation, and discuss the degree to which scintillation and profile evolution affect precision timing. This paper presents the data set as a basis for future, deeper studies.

\end{abstract}


\section{Introduction}

The International Pulsar Timing Array\footnote[1]{{\refc http://ipta.phys.wvu.edu}} (IPTA; {\refbfr \citealt{2010CQGra..27h4013H}}, \citealt{2013CQGra..30v4010M}) is a gravitational wave (GW) detector currently consisting of  $\sim50$ pulsars {\refbfr distributed} across the sky, monitored regularly by up to seven telescopes around the world: {\refbfr the Arecibo Observatory in the US, the Effelsberg radio telescope in Germany, the NRAO Green Bank Telescope (GBT) in the US, the Lovell radio telescope at Jodrell Bank Observatory in the UK, the Nan\c cay radio telescope in France, the Parkes telescope in Australia, and the Westerbork Synthesis Radio Telescope (WSRT) in the Netherlands}. Some of the pulsars in the IPTA have been precision-timed for a decade or more{\refbfr . These observations are performed by} the European Pulsar Timing Array (EPTA; \citealt{2013CQGra..30v4009K}), the North American Nanohertz Observatory for Gravitational Waves (NANOGrav; \citealt{2013CQGra..30v4008M}), and the Parkes Pulsar Timing Array (PPTA; \citealt{2013CQGra..30v4007H}, \citealt{2013PASA...30...17M}). The three collaborations combine their data as the IPTA. 

{\refr Pulsar timing compares times-of-arrival (TOAs) to those predicted
from a model that describes the pulsar's rotation, its binary motion,
the {\refxxxx interstellar medium (ISM)} between us and the pulsar, and the Earth's
motion in the Solar System. The measured TOAs are typically derived
{\refxxx from} pulsar profiles that have been averaged over the observation
duration; {\refxx and referenced against a high-precision frequency
standard at {\refxx the} observatories (typically hydrogen masers);}
which in turn is referenced to an international timing standard \citep{2004hpa..book.....L}.}
{\refbf {\refbfr {\refr If the resulting differences between measured and modelled
TOAs (the so-called ``timing residuals'') deviate significantly from zero, this indicates astrophysical
processes that are either not (or not completely) accounted for by the
timing model.} One {\refbfr possibility for such a process is long-period GWs {\refrr perturbing} the spacing between pulses as they propagate from a pulsar to the Earth. Obtaining accurate enough timing residuals {\refc to detect these GWs} requires repeated measurements over many years.} {\refbfr Sensitivity {\refc to GWs} increases as observation {\refc duration} grows, and the longest observation spans, as well as the red spectrum of the expected GWs, {\refx mean that the array is
most sensitive at a frequency of about 10\,yr} \citep{2013MNRAS.433L...1S}.} {\refg Individual TOAs are obtained by measuring the offset of emission beamed across the line-of-sight (LOS) at a given time from a template profile shape. Pulses can be averaged over a subintegration time, also known as ``folding'' according to a best-known pulse period.} The template profile is high-S/N and often averaged from long-term observations. From the radiometer equation {\refc relevant for pulsars} in \citet{2004hpa..book.....L} we have: {\refbfr
\begin{equation}
\textrm{S/N}\propto \frac{G\sqrt{t_{\textrm{int}}\Delta{f}}}{T_{\textrm{sys}}}
\end{equation}
in which S/N represents the integrated pulse S/N, $G$ the telescope gain, $t_{\textrm{int}}$ the pulse subintegration time, $\Delta{f}$ the bandwidth, and $T_{\textrm{sys}}$ the telescope's system temperature.} Thus subintegration time, bandwidth, and gain are all important observational parameters, with $G{\refxx /T_{\textrm{sys}}}$ most significantly impacting the reduction of radiometer noise, {\refg assuming we are comparing sensitivities for the same slice in frequency,} {\refc and given {\refrrrr that} the telescopes are all equipped with receivers having state-of-the-art $T_{\textrm{sys}}$ {\refxxxx levels}.}} {\refrrrr (Throughout this paper, pulse S/N will refer to the ratio of the peak pulse amplitude to the standard deviation of the mean-subtracted off-pulse amplitudes.)}

\footnotetext[16]{Kavli Institute for Astronomy and Astrophysics, Peking University, Beijing 100871, P. R. China}
\footnotetext[17]{Department of Physics and Astronomy, West Virginia Univ., Morgantown, WV 26506, USA}
\footnotetext[18]{CSIRO Astronomy \& Space Science, Australia Telescope National Facility, PO Box 76, Epping, NSW 1710, Australia}
\footnotetext[19]{Physics and Astronomy Department, University of New Mexico, 1919 Lomas Boulevard NE, Albuquerque, New Mexico 87131-0001, USA}
\footnotetext[20]{University of Bielefeld, Physics Department, D-33501 Bielefeld, Germany}
\footnotetext[21]{INAF-Osservatorio Astronomico di Cagliari, Via della Scienza 5, 09047 Selargius (CA), Italy}
\footnotetext[22]{School of Physics, Peking University, Beijing 100871, China}
\footnotetext[23]{Centre for Astrophysics \& Supercomputing, Swinburne University, Hawthorn, Victoria 3122, Australia}
\footnotetext[24]{ARC Centre of Excellence for All-Sky Astrophysics (CAASTRO), Swinburne University of Technology, Mail H30, PO Box 218, Hawthorn, VIC 3122, Australia}
\footnotetext[25]{Department of Physics, University of Vermont, Burlington, VT 05401, USA}

Pulsar timing arrays (PTAs) aim to detect perturbations due to GWs (\citealt{1978SvA....22...36S}, {\refbfr \citealt{1990ApJ...361..300F}}) in TOAs from millisecond pulsars (MSPs) on the order of 100\,ns {\refrrrr \citep{2004ApJ...606..799J}} after the TOAs are corrected for many other effects. These include terrestrial clock calibration, solar system ephemeris, {\refbfr variations in} dispersion measure (DM; proportional to the integrated LOS electron column density), proper motion, and position errors, all in the presence of noise due to other GWs at the source pulsars themselves. While the modeling uncertainties due to all the effects just listed are significant, GWs should perturb TOAs in a correlated way across the sky as a function of angle-of-separation between pulsars \citep{1983ApJ...265L..39H}.} This correlation makes the detection criterion {\refbfr less sensitive to any systematic errors in the TOAs or in the timing model for any one pulsar.} Detectable strains (spatial strains due to GWs; $h$) are {\refg expected to be} on the order of {\refbfr $h\sim10^{-15}$ at nHz frequencies \citep{2013MNRAS.433L...1S}.} Plausible sources {\refbfr producing} GW strains in the PTA {\refbfr frequency range} include: a stochastic background of GWs (\citealt{1979ApJ...234.1100D}, \citealt{1983ApJ...265L..39H}) due to merging supermassive black hole binaries (SMBHBs), continuous wave sources from individual SMBHBs in $z<1.5$ galaxies \citep{2009MNRAS.394.2255S}, bursts on timescales of months from SMBHBs in highly elliptical orbits \citep{2010ApJ...718.1400F}, cosmic strings {\refbfr (\citealt{1979JETPL..30..682S}, \citealt{2013ApJ...764..108S})}, phase transitions in the early universe \citep{PhysRevD.82.063511}, and relic GWs from the era of inflation \citep{2005PhyU...48.1235G}. {\refbfr Additionally, PTAs make possible the detection of GW bursts-with-memory, signals that are anticipated from events such as the final merger of SMBHBs and potentially from exotic phenomena at extremely high redshift (\citealt{2010MNRAS.401.2372V}, \citealt{2012ApJ...752...54C}, \citealt{2014ApJ...788..141M})}.

{\refbfr Through the IPTA consortium, all three PTAs (NANOGrav, the EPTA, and the PPTA) share timing data from their seven different observatories.} The seven telescopes have different receivers, backend instruments, sensitivities, and radio frequency interference (RFI) environments, and have been observing their selected sets of pulsars for a range of epochs. Each telescope also has a history of regularly improving {\refbfr instrumentation}, and thus TOAs obtained at later times are often of a much higher quality than those from earlier times. This trend is helpful for timing precision, but a wider bandwidth may require a more complicated frequency-dependent pulse profile model, due to frequency-dependent pulse shapes {\refbfr (\citealt{Liu14}, \citealt{2014ApJ...790...93P})}. Differences amongst PTAs include the number of standard observing frequencies and the methods for modeling DM variations. {\refbf Fortunately, many of these difficulties in data combination are not insurmountable, and tremendous progress has already been made (see \citealt{2013CQGra..30v4010M}).} {\refbfr The benefits of such a combination are many, and include an improved cadence, cross checks, better frequency coverage, and more pulsars {\refg correlated} across the sky.}

Apart from the need to combine data from many telescopes, there is also the need to better understand what {\refbfr might intrinsically limit timing quality. PSR~J1713+0747 \citep{1993ApJ...410L..91F} is regularly observed by all IPTA telescopes, and provides much of the sensitivity for GW upper limit calculation \citep{2014ApJ...794..141A} with a timing stability of $\sim100$\,ns on timescales of five years or more \citep{2009MNRAS.400..951V}.} In contrast, the first MSP discovered, PSR~B1937+21, is well known to be extremely stable on the order of weeks to months, {\refbfr but its residuals show a significant red noise} power spectrum visible on timescales of years \citep{1994ApJ...428..713K}. {\refbfb As we design larger telescopes and observing programs, it is imperative that we know the fundamental limits of timing precision, i.e. at what point additional gain, observing time, {\refg or} bandwidth will not increase our timing precision.}

Upper limit papers such as those by \citet{2013Sci...342..334S}, \citet{2011MNRAS.414.3117V}, and \citet{2013ApJ...762...94D} {\refbfr have all calculated GW limits based on TOAs} over 5+ years. {\refbf If the observation duration at a single epoch were increased from the typical subintegration time} by a factor of X, the timing precision (in the absence of other limiting effects) would be naively expected to improve by $\sqrt{\mathrm{X}}$ {\refr as in Equation 1}. This is simply due to the fact that the number of pulses collected would increase, if TOAs (in the absence of GWs) can be fitted to standard timing models such that the residuals are white noise, assuming there are no significant pulse shape changes {\refbfr between observation epochs}. 

Many properties of a pulsar along its particular LOS are not precisely predictable: DM variations, interstellar scintillation (ISS), scattering {\refbf variations}, and low-level glitches, to name a few. Glitches are not observed in PSR~J1713+0747 {\refr or in MSPs in general \citep{2011MNRAS.414.1679E}}, though they may be present at low amplitudes in many pulsars (see,
however, \citealt{2014MNRAS.440.2755E}) and {\refr therefore may} act as a limiting factor in searches for GWs; they may be especially problematic sources of noise in searches for GW bursts-with-memory.  {\refxxx M28A is another exception \citep{2004ApJ...612L.125C}, though admittedly this may be because it is particularly young for an MSP.}

{\refbfr Pulse phase jitter, which is independent of radio frequency, is also a limiting factor for pulsar timing.} Jitter, also known as pulse-to-pulse modulation, was first described in \citet{1985ApJS...59..343C} for canonical pulsars, established for PSR~B1937+21 by \citet{1990ApJ...349..245C}, and {\refrr more recently measured in PSR~J1713+0747 in \citet{2014MNRAS.443.1463S} and \citet{2012ApJ...761...64S}}. {\refg The term refers to the distribution of arrival times of single pulses about the the peak of the averaged template pulse, which have slight offsets in pulse phase.} {\refbfr Neither increasing telescope size nor bandwidth} eliminates the presence of TOA jitter errors, jitter being both broadband and independent of pulse profile {\refbfr S/N}. 
{\refg {\refxxxx Generally,} the only way to reduce jitter-induced white noise
in most pulsars is to increase the observation duration
\citep{2004NewAR..48.1413C}, though there are exceptions; see \citet{2013MNRAS.430..416O} and \citet{2011MNRAS.418.1258O} for an example of a jitter mitigation technique
on {\refr PSR~J0437$-$4715.}

These considerations motivate observations of PSR~J1713+0747 for 24 continuous hours, using nine radio telescopes: the seven IPTA telescopes along with LOFAR (LOw Frequency ARray, \citealt{2013A&A...556A...2V}, {\refbfr \citealt{2011A&A...530A..80S}}) in the Netherlands and the GMRT (Giant Meterwave Radio Telescope) in India. See Table~1 for details of the allotted frequencies for each telescope. {\refbfr The duration of 24\,hr was chosen because MSP timing has been explored at the hour and week timescales, but not in the intermediate regime.} The inclusion of LOFAR provided an ultra-low frequency {\refbfr (110 -- 190\,MHz)} component to the observation, sampling a frequency range that features prominent effects from the {\refbfr interstellar medium}. {\refbfr Observing at} {\refbf L-band/1.4\,GHz is ideal for {\refxxxx studying the timing properties of} this particular pulsar, being {\refbfg reasonably bright} given its {\refbfr flux} spectrum with a power-law of slope of --1.7 \citep{2001A&A...368..230K}}, but not significantly affected by red noise in the timing residuals due to the ISM \citep{2013MNRAS.429.2161K}. The GMRT filled in the time coverage gap between Parkes and the European telescopes, enabling a continuous 24\,hr of observing. 

\begin{deluxetable*}{cccccccccc}[l]
\tablecolumns{10}
\tablecaption{Observing Schedule and Parameters}
\tablehead{
\colhead{Telescope} & \colhead{Obs.} & \colhead{Start UT} & \colhead{End UT} & \colhead{Min.} & \colhead{Cent.} & \colhead{Bandwidth} & \colhead{{\refbfr Min. Channel}} & \colhead{Pulse Profile} & \colhead{Time to}\\
\colhead{} & \colhead{Mode} & \colhead{} & \colhead{} & \colhead{Subint.} & \colhead{Freq.} & \colhead{(MHz)} & \colhead{{\refbfb Width}} & \colhead{S/N$^{\textrm{a}}$ for Obs.} & \colhead{Stated}\\
\colhead{} & \colhead{} & \colhead{} & \colhead{} & \colhead{Length (s)} & \colhead{(MHz)} & \colhead{} & \colhead{Possible (MHz)} & \colhead{Duration$^{\textrm{b}}$} & \colhead{S/N (hr)$^{\textrm{c}}$}
}
\startdata
Arecibo & Int$^{\textrm{d}}$ & 23-JUN 03:17 & 23-JUN 04:44 & - & 1382 & 800 & 6.25 & {\refxxxx 3138} & 1.45\\
Effelsberg & Fold$^{\textrm{e}}$ & 22-JUN 18:26 & 23-JUN 03:40 & 10 & 1348 & 200 & 1.56 & {\refxxxx 473} & 7.20\\
GBT & Int & 22-JUN 00:56 & 23-JUN 10:15 & - & 1497 & 800 & 6.25 & 2200 & 8.95\\
GMRT & CF$^{\textrm{f}}$ & 22-JUN 13:38 & 22-JUN 21:58 & 60 & 1387 & 33.3 & 1.04 & 80 & 7.63\\
Lovell & Fold & 22-JUN 18:04 & 23-JUN 05:34 & 10 & 1532 & 400 & 0.25 & {\refxxxx 404} & 9.85\\
Nan\c cay & Fold & 22-JUN 22:33 & 22-JUN 23:30 & 60 & 1524 & 512 & 16 & 125 & 0.95\\
Parkes & Fold$^{\textrm{g}}$  & 22-JUN 10:20 & 22-JUN 16:20 & 60 & 1369 & 256 & 0.25 & {\refxxxx 344} & 6.00\\
\hline
LOFAR & CF & 22-JUN 18:11 & 23-JUN 03:00 & {\refxxx 5} & 148.9 & 78.1 & {\refxxx 0.195} & 8 & 8.82\\
Westerbork & CF & 22-JUN 21:46 & 23-JUN 04:39 & 10 & 345$^{\textrm{h}}$ & 80 & 8.75 & {\refxxxx 27} & 4.95
\enddata
\footnotetext{All S/N values are scaled to 512 phase bins. {\refc These S/N values are affected by both scintillation and observation length as well as the\\ telescope parameters.}}
\footnotetext{{\refc The duration here refers to the duration of the folding portions of a telescope's observation run only.}}
\footnotetext{{\refr In addition to observation time, {\refrrr scintillation} also significantly influenced these S/N values.}}
\footnotetext{{\refbfr Intensity recording mode (non-folding). Single pulses are the minimum subintegration time.}}
\footnotetext{{\refbfb Normal pulse folding mode, using coherent dedispersion.}}
\footnotetext{{\refbfb {\refrrrr Offline} coherent filterbank mode. For these multi-antenna telescopes {\refrrr acting as a phased array}, folding and coherent dedispersion is applied offline.}}
\footnotetext{{\refbfb The Parkes DFB3/4 backend, one of three backends used in parallel, does not apply coherent dedispersion online.}}
\footnotetext{{\refc Observations alternated between 345\,MHz and 1398\,MHz as center frequencies. Refer to Figure~1 and Table~2 for details.}}
 \end{deluxetable*}
\begin{deluxetable*}{cccccc}
\tablecolumns{6}
\tablecaption{Baseband Observing Parameters}
\tablehead{
\colhead{Telescope} & \colhead{Obs. Mode} & \colhead{Start UT} & \colhead{End UT} &  \colhead{Cent. Freq.} & \colhead{Bandwidth}\\
\colhead{} & \colhead{} & \colhead{} & \colhead{} & \colhead{(MHz)} & \colhead{(MHz)}\\
}
\startdata
Arecibo & BB\footnote{Baseband voltage recording mode} & 23-JUN 02:30 & 23-JUN 03:00 & 1378 & 200 \\
Effelsberg 	&	 BB 	&	22-JUN 22:15	&	 22-JUN 23:45  	&	1396	&	128	\\
	&	 BB 	&	23-JUN 02:05	&	 23-JUN 02:50  	&	1396	&	128	\\
GBT 	&	 BB 	&	 22-JUN 22:15 	&	 22-JUN 23:15 	&	1378	&	200	\\
 	&	 BB 	&	 23-JUN 01:50 	&	 23-JUN 02:52 	&	1378	&	200	\\
GMRT 	&	 BB 	&	 22-JUN 22:24 	&	23-JUN 00:00	&	1387	&	33.3		\\
Lovell 	&	 BB 	&	 22-JUN 22:14 	&	 22-JUN 23:44 	&	1396	&	128	\\		
 	&	 BB 	&	 23-JUN 01:47 	&	 23-JUN 02:56 	&	1396	&	128		\\
Parkes 	&	 BB 	&	 22-JUN 10:20 	&	 22-JUN 16:20 	&	1369	&	256	\\
Westerbork 	&	 BB 	&	 22-JUN 22:16 	&	 22-JUN 23:45 	&	1398	&	128	\\
 	&	 BB 	&	 23-JUN 02:05 	&	 23-JUN 02:51 	&	1398	&	128	
\enddata
 \end{deluxetable*}
\begin{figure*}
\centerline{\includegraphics[width=2.6\columnwidth]{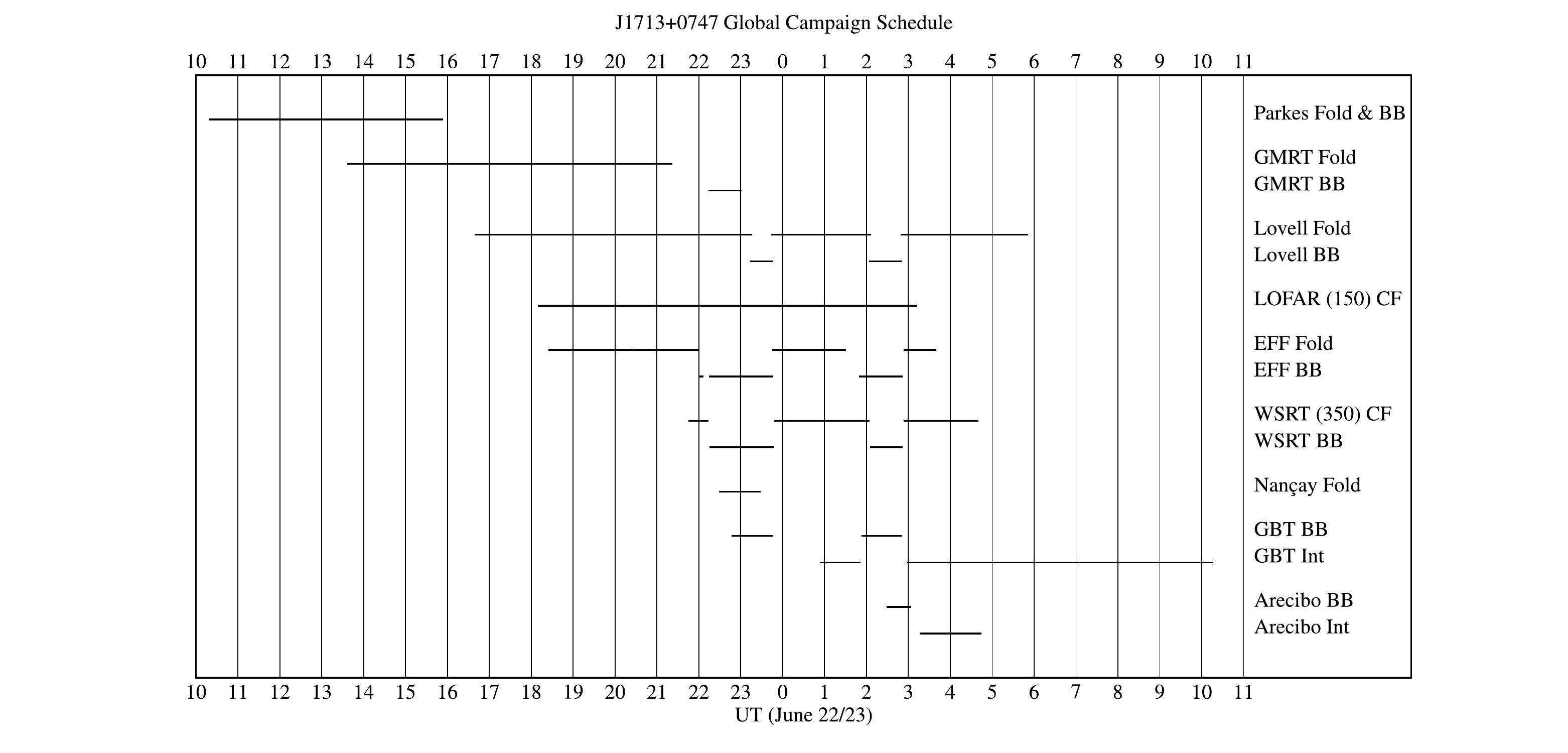}}
\figcaption{{\refbfr Timeline of the global observation of PSR~J1713+0747, showing the various telescopes and observing modes.} Here, ``Fold'' refers to ordinary pulse folding with coherent dedispersion. ``BB'' refers to baseband mode {\refc at L-band/1.4\,GHz}, recording complex voltages without folding. ``Int'' refers to intensity integrations, also known as ``coherent search mode'', which is similar to baseband mode except recording an intensity rather than full voltage information. ``CF'' refers to the {\refrrrr offline} coherent filterbank mode used at LOFAR and WSRT, in which the telescope recorded a
coherent sum (tied-array beam) of all the antennas, which was written out as complex voltages and then coherently dedispersed and folded offline.}
\end{figure*}

Another major science goal of the global PSR~J1713+0747 observation relates to the LEAP project (Large European Array for Pulsars; {\refrr \citealt{Bas14} in preparation}, \citealt{2013CQGra..30v4009K}), which uses the Effelsberg, Nan\c cay, Lovell, WSRT, and Sardinia radio telescopes as a phased array, together as sensitive as Arecibo, with a comparable total collecting area ($3\times10^4$\,m$^2$){\refbf, but with a much greater {\refrrrrr observable} declination range than Arecibo}. As PTAs advance, this configuration may prove to be critical for the detection of GWs. The {\refx 24-hr} global observation of PSR~J1713+0747 helps the LEAP effort by adding {\refbfb three more telescopes -- Arecibo itself, as well as the Green Bank Telescope and the GMRT. The present dataset therefore opens up the possibility of experimenting with a telescope having over} twice the collecting area of Arecibo alone. The combined effort will be referred to in the present paper as GiantLEAP.

This dataset also represents a unique opportunity to measure clock offsets between telescopes and how they vary across {\refbf overlapping} time intervals. When combining TOAs from different telescopes (or when backends change on a single telescope) an offset or ``jump'' is needed. Such a jump can be due to {\refbfr delays in the backends themselves, such as cable delays,} conspiring with factors that are difficult to quantify individually (see \citealt{2013CQGra..30v4001L} and \citealt{2013CQGra..30v4009K} for further explanation). The jumps can then be quantified by fitting for one arbitrary timing offset per telescope/backend pair per frequency {\refbfr only in the overlapping region}, such that the rms of the combined dataset is minimized. Simultaneous observations -- the longer the better -- provide an opportunity to measure such offsets and their drifts with high accuracy.

Timing stability on the {\refg $\sim24$\,hr} timescale can also be {\refbfr quantified using} the Allan variance of the residuals. {\refbfr The Allan variance was originally used to quantify the stability of} atomic clocks (see \citealt{1997A&A...326..924M} for details). The present dataset allows us to evaluate the Allan variance for clock frequencies of 10$^{-2}$\,Hz all the way into {\refbfr frequencies corresponding to} the five or even twenty year datasets (\citealt{Zhu15}, in preparation) that exist for PSR~J1713+0747.

Yet another goal of the PSR~J1713+0747 24-hour global campaign is to assess a noise floor. {\refbfr How does rms precision increase as longer timespans of data are analyzed?} Does the improvement ``bottom out'' or continue indefinitely with the number of collected pulses on these timescales?

{\refbfr We intend the present paper to be a description of the data set itself as well as being an introduction to a series of papers}, given the size of the dataset and the large number of science goals. Here we present some first science results, and intend to expand upon them and address other topics in later papers. {\refbfr In Section~2, we describe the observation and the resulting dataset in detail, supplemented by the Appendix. In Section~3, we explore a number of first results emerging from the all-telescope analysis. Section~4 shows some general first results on the timing error budget, and finally Section~5 mentions some future paper topics based on the data.}

\section{Description of the Data}
The observations were conducted on 22 Jun 2013 (MJD 56465 -- 56466) starting with the Parkes Telescope and progressing through the other eight telescopes for as much time as possible between rise and set. The observation timeline can be seen in Figure~1. {\refbf The time of year was such that local midnight roughly corresponded to the middle of each telescope's observation.} PSR~J1713+0747 was approximately 146$^\textrm{o}$ from the Sun, minimizing the possibility of any solar effects on the data \citep{2007ApJ...671..907Y}.
\begin{figure*}
\includegraphics[width=2.2\columnwidth]{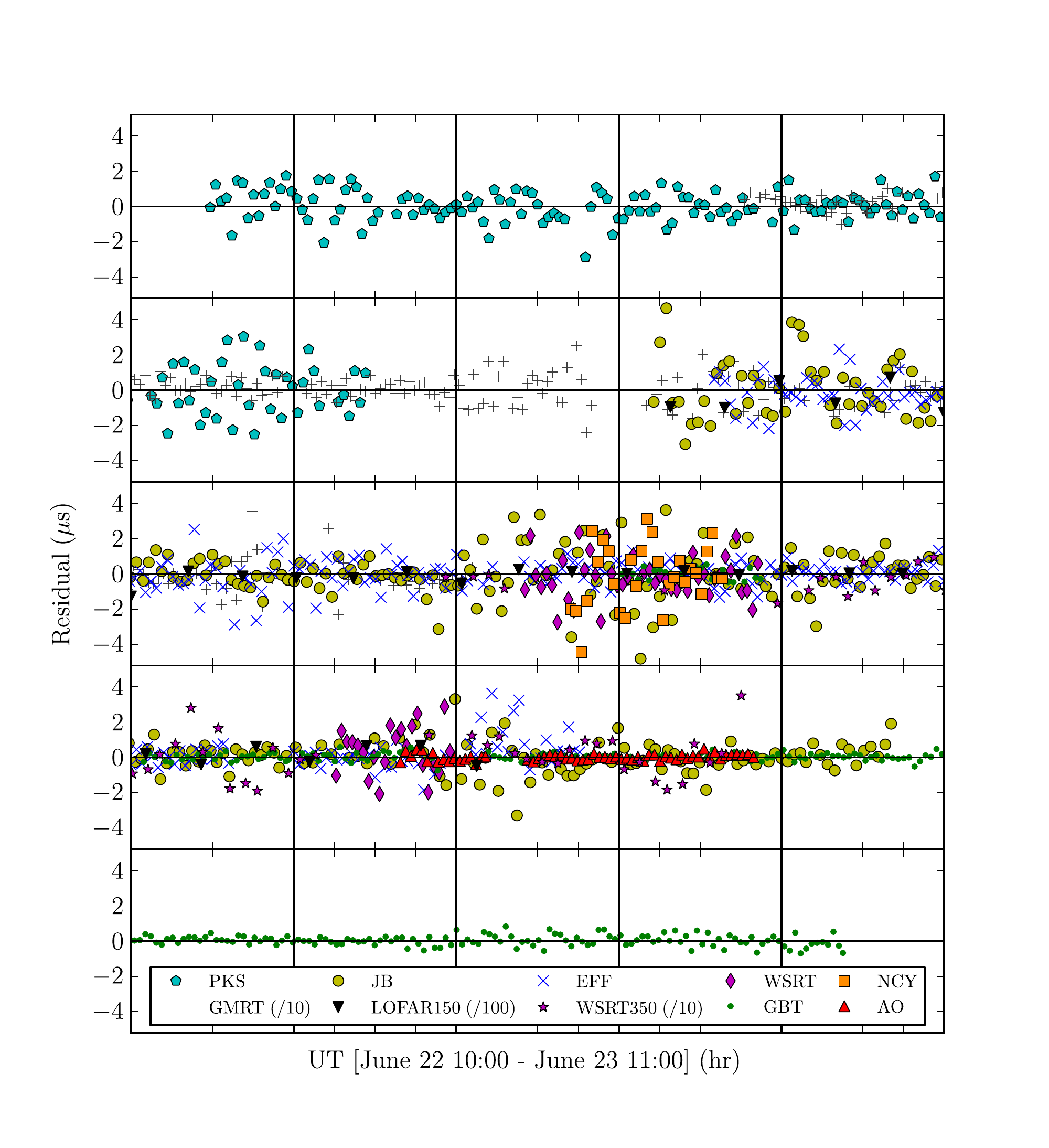}
\figcaption{24-Hour Timing Residuals: timing residuals for {\refbfg the nine} telescopes as a function of time. {\refrrrr Values shown are for L-band/1.4\,GHz observations unless otherwise noted.} {\refx Uncertainties} on each residual are not shown here in order to maintain clarity. In all cases, the fitting error on individual residuals is on the order of the scatter of all residuals shown for a particular telescope. All residual values are for 120\,s integrations, except LOFAR which is for 20\,min. The increase in residual values in the third and fourth rows from the top is due to most telescopes having switched to the lower bandwidth baseband mode. Residual values from the GMRT, WSRT-350\,MHz, and LOFAR-150\,MHz are scaled down by factors of 10, 10, and 100 respectively {\refbfr in order to show all residuals in a single panel}.}
\end{figure*}
\begin{sidewaysfigure*}
  \centering
  \centerline{\includegraphics[trim=-3.5cm 0cm -5.5cm -14cm,clip=false,width=1.3\columnwidth]{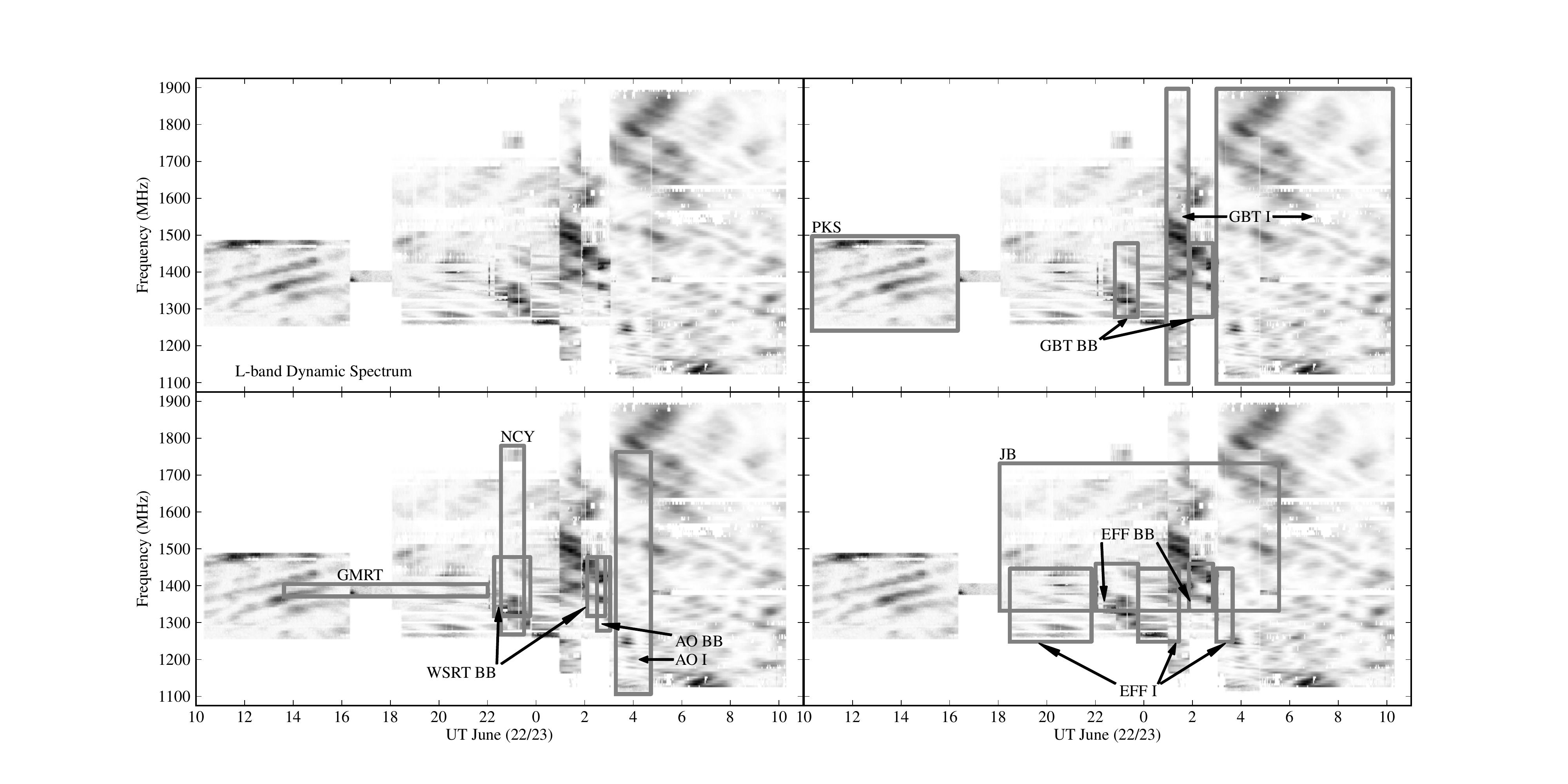}}
\figcaption{Dynamic Spectrum: L-band/1.4\,GHz dynamic spectrum for 8 of the 9 telescopes (excluding LOFAR at 150\,MHz and WSRT at 350\,MHz). Note the narrowing of the scintillation bandwidth as a function of frequency, along with a decrease in scintillation timescale with decrease in frequency. Scintillation patterns are shown to overlap well between telescopes, thus establishing {\refxxxx (in Section~3.3)} a typical spatial scale for the scintillation pattern. A cross-hatched pattern is clear, which is usually only observable at low frequencies, {\refbfr but observable here because of the large bandwidth and {\refx 24-hr} observation time}. {\refc When observations overlap, the regions of the dynamic spectrum shown} are from telescopes with the highest sensitivity per patch $dA = d\nu{dt}$.}
\end{sidewaysfigure*}

As Figure~1 shows, there were different modes used amongst the telescopes -- ordinary pulse folding, {\refbfb baseband mode, {\refg coherent filterbank mode {\refxx (formed offline)}}, and non-folded intensity integrations (also known as ``coherent search mode''). For ordinary pulse folding, coherent dedispersion is applied in real time \citep{1975mcpr...14...55H}, correcting for pulse delays due to dispersion in the ISM across few-MHz channels {\refbfr (with the exception of the DFB backends at Parkes in which dedispersion is applied after the fact -- see Appendix)}. {\refbfr All the non-baseband data presented here have coherent dedispersion applied.} {\refg Observing in baseband mode affords a number of advantages.} For the science goal of expanding LEAP with Arecibo and GBT, baseband recording is a {\refbfr requirement of any phased array formed offline}. It also allows us to evaluate how cyclic spectroscopy (\citealt{2011MNRAS.416.2821D}, \citealt{2013ApJ...779...99W}, \citealt{2013CQGra..30v4006S}) might improve the quality of some of the dataset as a method of obtaining a significantly higher frequency resolution (\citealt{2014ApJ...790L..22A}, \citealt{Jon15}, in preparation). 
{\refrr The baseband sessions were
driven by when the transit telescopes (Arecibo and Nan\c cay) could
observe the pulsar. Hence, the first baseband session was when the
source transited at Nan\c cay, when baseband data was obtained at GMRT,
Lovell, Effelsberg, Westerbork, GBT and Nan\c cay, while the second
session was when the source transited at Arecibo, and baseband data
was obtained with the Lovell, Effelsberg, Westerbork, GBT and
Arecibo telescopes.}
Non-folded intensity integrating is similar to baseband mode, except that intensities rather than signed voltages are recorded, without phase information, resulting in a more manageable data size. Both} baseband data and non-folded intensity recording yield single pulse information. {\refbfr Otherwise, single pulses are not recoverable due to the folding process.}

{\refxxxx Baseband was not, in general, the default observing mode.} Taking baseband data for the entire rise-to-set time at each telescope would be cumbersome in terms of data volume (which {\refc as it is} approached {\refc 60\,TB} total largely due to the baseband component) and in some cases would also limit bandwidth. For example, in the case of the GUPPI/PUPPI backends (Green Bank / Puerto Rican Ultimate Pulsar Processing Instrument; \citealt{2008SPIE.7019E..45D}), baseband recording {\refbfr can only be conducted over} 200\,MHz of bandwidth, {\refbfr as opposed to the 800\,MHz available for the folding and intensity integration modes}. The overlapping baseband portion of the observation for GiantLEAP consisted of two baseband allotments (see Figure~1). Arecibo could only participate in one of these sessions due to the limited range in zenith angle. Parkes observed with baseband data taking in parallel for its entire observation duration.

Reduced profiles are in the \textsc{psrfits} format processed with the \textsc{psrchive} and \textsc{dspsr} software packages (\citealt{2004PASA...21..302H}, \citealt{2012AR&T....9..237V}), with final timing residuals determined by the \textsc{tempo2} software package \citep{2006MNRAS.369..655H}. {\refbfr Residuals are generated with a common \textsc{tempo2} parameter file, after having produced TOAs with \textsc{psrchive} based on} standard observatory-specific template pulse profiles, {\refr noise-reduced using \textsc{psrchive} when necessary.} Different profiles are used because of local bandpass, frequency range, and calibration differences. We create TOAs {\refbf from 120\,s fiducial subintegrations} in common between L-band/1.4\,GHz telescopes. This choice of subintegration time was a compromise between, on the one hand, {\refq having a sufficient number of TOAs in order to probe timing precision on long timescales (see Section~3.1),} and on the other, having {\refbfb a minimum TOA S/N of approximately 1 across all L-band/1.4\,GHz} telescopes {\refq (see Section~3.2 for an application to measuring jitter)}. Further details about specific telescopes can be found in the Appendix and in Table~1, in which the S/N of each telescope's 120\,s TOAs can be found, with baseband details in Table~2. As can be seen in Table~1, the {\refbfr possible minimum subintegration length is less than 120\,s for most telescopes, providing flexibility for future studies.} Slots in which the table is blank indicate intensity integrations, which can be integrated to any time greater than $2/\Delta{\nu}$, where $\Delta{\nu}$ is the {\refg channel width}. {\refbfb WSRT, GMRT, and LOFAR, being multi-antenna telescopes, observed in a tied-array mode (formed online) which dumped complex voltage data to be coherently dedispersed and folded using \textsc{dspsr}. We refer to this as the ``coherent filterbank'' mode because a filterbank is formed offline, except for GMRT, for which we still refer to the modes used as folding and baseband, given the additional optimization employed (see Appendix for details).

We remove RFI automatically in post-processing using \textsc{psrchive}. The program defines an off-pulse window by iteratively smoothing the profile and finding the minimum. Using the maximum minus minimum intensity values defined in this window, we apply a median filter to identify RFI spikes and flag spikes at greater than 4\,$\sigma$ {\refbfr from half the intensity difference}. Any {\refg remaining} particularly noisy frequency channels or time integrations are manually removed after visual inspection. {\refbfr At Lovell, RFI is excised in real time. For the other telescopes, post-facto algorithms are used -- one that excises ``rows'' and ``columns'' of RFI {\refbfb in frequency and time}, and another that excises RFI within the pulse profile itself according to bad phase bins.}

The residuals shown in Figure~2 were generated by folding the respective data sets from each individual telescope with a parameter file that was {\refbf created from 21 years of PSR~J1713+0747 data (\citealt{Zhu15}, in preparation), all with the same Bureau International des Poids et Mesures (BIPM) correction table to the global atomic timescale, TT(BIPM2012), with the 2013 extrapolation, and the Jet Propulsion Laboratory DE421 planetary ephemeris. The DM ($15.99113\pm0.00001$\,pc\,cm$^{-3}$)} is the only fitted {\refg astrophysical} parameter determined from the present observation across all {\refbfr L-band/1.4\,GHz} telescopes simultaneously. {\refbfb Relative offsets were also fitted between telescopes. These are simply free parameters that align the residuals and do not represent absolute clock offsets, and are not shown here for this reason. {\refc The remainder of the parameters were held fixed at the \citealt{Zhu15} (in preparation) values} (see Table~3 in the Appendix for further details).} The timing {\refbfr residuals at other, higher subintegration times} are computed from this base set of residuals. To test {\refbfr that residuals can be averaged down without a loss in modeled timing precision}, we also generated a set of 10\,s subintegration time TOAs with {\refrrrr \textsc{psrchive}} from the GBT, and found that simply averaging the 10\,s residuals produced new residuals with rms values different from the 120\,s \textsc{tempo2} residuals at $\ll$\,1\,$\sigma$. {\refx Therefore, we can create 10\,s residuals, average them, and obtain nearly identical 120\,s residuals to those resulting from 120\,s TOAs.}

The eight telescopes at L-band/1.4\,GHz saw a changing spectrum due to interstellar scintillation, as shown {\refbfr in the dynamic spectrum, or plot of pulsar intensity vs. time and frequency, in Figure~3. Scintillation is due to the scattering and refraction of pulsed emission in the ionized ISM, }{\refr and can significantly change the pulse profile S/N as can be seen in Table~1. {\refxx Figure~3} shows that bright scintles contribute to the high S/N of telescopes after about 23-JUN 00:00. }{\refc The image was created by fitting the telescopes' templates via matched filtering for a given telescope with a given profile} $P(\nu,t)$ and reporting the amplitude {\refbfr with the off-pulse mean subtracted. Amplitudes are converted to corresponding intensities depending on each telescope's calibration data and then scaled to an identical color scaling for Figure~3.} {\refq Dynamic spectra from different telescopes were scaled empirically to match them in Figure~3, lacking absolute calibration for some telescopes, {\refr but using the noise diode calibrations employed} (see Appendix for details).} A cross-hatched pattern is clear. Such a pattern is usually only observable at low frequencies due to the small scintle size in frequency and time even for modest bandwidths \citep{1970MNRAS.150...67R}. Note the narrowing of the scintillation bandwidth ($\Delta_\nu$, the typical scintle width in frequency) as a function of frequency, along with a decrease in scintillation timescale ($\Delta_t$, the typical timescale for scintillation) with decrease in frequency. Scintillation patterns are shown to overlap well between telescopes, thus establishing a typical spatial scale of a waveform -- see Section~3.3. {\refc In some cases there appear to be deviations between the scintillation patterns seen at different telescopes, but this always
corresponds to times when the source was close to rising or setting at one of the telescopes involved.}

Narrowband template fitting \citep{1992RSPTA.341..117T} assumes a relatively constant profile with frequency. In addition to distorting the pulse phase, merely averaging across frequency would result in drifting residuals with a non-white appearance \citep{1970PhDT.........8C}, {\refbfr due to the intrinsic profile evolution acting in combination with ISS}. Profile shape changes with frequency are present in all canonical pulsars -- see \citet{1986ApJ...311..684H} for multifrequency observations on many pulsars, and \citet{2013A&A...552A..61H} which uses observations from LOFAR and other telescopes. Similar shape changes have also been found in MSPs \citep{1999ApJ...526..957K}, including PSR~J1713+0747.

{\refrrrr Figure~4 shows the presence of profile evolution with frequency in this observation's GBT data. Starting with 8\,hrs of GBT data, we use the fiducial subintegration length of 120\,s and a subband size of 50\,MHz. We sum profiles in time to get 16 profiles as a function of observing frequency and phase, $P(\nu,\phi)$. These profiles are dedispersed using the best fit, L-band/1.4\,GHz DM. For each $P(\nu,\phi)$, we fit a NANOGrav standard template $T(\phi)$ to the data profile $P(\nu,\phi)$ to find the best-fit phase offset $\delta\phi(\nu)$ and amplitude $A(\nu)$. We create difference profiles by shifting and scaling $T(\phi)$ using the best fit phase offset $\delta\phi(\nu)$ and amplitude $A(\nu)$ for each frequency and then subtracting as $D(\nu,\phi)  = P(\nu,\phi)- A(\nu)T(\phi-\delta\phi(\nu))$, where $D(\nu,\phi)$ are the difference profiles. These are plotted in the main panel of Figure~4. The right panel shows {\refbfr the timing offsets as a function of frequency, $\delta\phi(\nu)$, with the 1422\,MHz profile set to zero offset because the template most closely resembles these data in the center of the band. These timing offsets will be dependent on the value of our measured DM, which in turn is dependent on the frequency dependent (FD) model parameters {\refg (\citealt{Arz15}, in preparation)} used {\refrrrr in \textsc{tempo2}}. {\refx The FD parameters correspond to the coefficients of polynomials of the logarithm of radio frequency that show the TOA shift due to profile evolution.} If the profile evolution within a subband is small and if the FD model parameters quantify the offsets well when each subband is independently used to create a set of timing residuals, then the weighted broadband residuals should be consistent with the white noise expected from {\refrrr scintillation}. As Figure~2 shows for all the telescopes, the residuals are qualitatively white noise-like in character. The broadband weighting, then, appears to correctly take the profile evolution / scintillation interaction into account. {\refg The profile evolution shown with frequency in Figure~4 is likely to be intrinsic and not an instrumental artifact because the equivalent Arecibo plot {\refc (i.e. using a different receiver at a different telescope)} is nearly identical across the same bandwidth. {\refbf A more detailed analysis of the observed profile frequency evolution is a subject of future work.}}}

We address this profile-evolution problem for all telescopes with bandwidths of 100\,MHz or more (that is, all telescopes except the GMRT) by computing TOAs for multiple narrow frequency channels, using \textsc{psrchive} as described above. For the GBT and for Arecibo, the band is divided into {\refxxxx 16 bins of 50\,MHz each}. The data from the other telescopes are split into subbands in similar fashion. We then obtain sets of narrowband timing residuals using the FD parameters in \textsc{tempo2}. {\refg The four best-fit FD parameters can be found in Table~3 {\refxx in the Appendix}, representing third-order polynomial coefficients starting with the lowest order first.} We then perform a weighted mean of these values in order to obtain the broadband residuals. For all telescopes we use the FD parameters {\refbfr with} \textsc{tempo2} independently computed from \citet{Zhu15}.

{\refbfr Intrinsic pulse profile evolution is thought to arise from varying offsets between the emission region and the surface of the neutron star, with higher frequency emission being produced closer to the surface} (see \citealt{2013CQGra..30v4002C} for a more detailed discussion). {\refbfr If a profile at a high narrowband frequency differs significantly from} a profile at a low narrowband frequency, then any frequency-dependent pulse shape changes will be highly covariant with the DM measurements at each epoch. {\refbfr Multi-frequency timing minimizes such covariances. {\refbf The timing offsets due to intrinsic pulse shape changes with radio frequency are constant in time.} Effects due to interstellar scintillation and scattering will depend on time, however. The former produces a varying S/N across the band {\refbfr due to scintillation} that changes the relative weighting of each subband as part of the final TOA; the latter broadens the pulse, resulting in a scattering delay. {\refbfr (The L-band/1.4\,GHz frequency is chosen, for the present dataset and for most standard timing observations, so that scatter broadening is minimal.)}}}

\section{Analysis of the Multi-Telescope Data}  

The following {\refg assumptions and terminology} will be used throughout this section:

The rms of the timing residuals over the total time span is consistent with the errors expected from {\refxxxx a} finite S/N ratio and from single-pulse stochasticity that is intrinsic to the pulsar. {\refxxxx From \citet{2010arXiv1010.3785C}, finite S/Ns yield an {\refbf approximate} template fitting error of}
\begin{eqnarray}
\sigma_{\mathrm{S/N}} = \frac{W_{\mathrm{eff}}}{S(N_{\phi})\sqrt{N_{\phi}}},
\label{eq:eq1}
\end{eqnarray}
{\refxxxx where $S(N_{\phi})$ is the S/N of the pulse profile (peak to rms off-pulse) that has $N_{\phi}$ phase bins and $W_{\mathrm{eff}}$ is the effective pulse width. \citet{2010arXiv1010.3785C} give an expression for $W_{\mathrm{eff}}$ that we use for PSR~J1713+0747, yielding 0.54\,ms.}

Here, $N_{\phi}$ is included because for the GMRT, $N_{\phi}$ was 64, while for the other telescopes, $N_{\phi}$ was 512, and this difference has been noted in all relevant calculations. {\refbfr These values for $N_{\phi}$ are chosen such that $N_{\phi}$ is small enough to afford sub-$\mu$s timing precision {\refr at some telescopes}, while at the same time, producing a $S(N_{\phi})$ {\refg small enough such that the pulse peak measurement is reasonably accurate.}} {\refbfg When we use $\sigma_{\mathcal{R}}$, it will refer to the total residual rms, whether template fitting error, jitter, radiometer white noise, or white noise due to the ISM.}

\subsection{Timing Residual Precision vs. Integration Time}

In Figure~5 we show the logarithmic change of the {\refr L-band/1.4\,GHz} {\refbfr TOA residual rms, a proxy for timing precision,} as both a function of subintegration time $T$ and of the corresponding number of pulses $N$. The time per TOA is plotted on the abscissa, and the corresponding number of pulses for each residual subintegration time {\refbfr is also shown.} {\refbf The ordinate shows the rms of the residuals within the entire observation time of the telescope. For this reason, the data points for Arecibo do not extend to as long of a timescale as the other telescopes despite the high sensitivity.} We start from TOAs from the base subintegration time of 120\,s, and integrate down {\refbf (i.e. to larger subintegrations)} for each successive step. {\refbf We show this function for five of the L-band/1.4\,GHz telescopes, choosing the maximum integration time at each telescope which corresponds to {\refg at least eight subintegrations} in order to ensure that small-number statistics (due to having only  a few long-subintegration residuals) are not important. Nan\c cay and WSRT are not shown because of their short, non-contiguous observing times. GMRT's residuals did not probe {\refbfr small} values of $N$ and are not shown.} Error bars are 1\,$\sigma$ and are simply the {\refbfr standard error of the scattered subintegration rms values} in a block of TOAs.} For reference, the expected $1/\sqrt{N}$ slope is plotted. {\refbf Each successive data point is not independent of the data points in Figure~5 for small values of $N$.} In all telescopes, there are no significant deviations from this simple improvement in timing rms with number of pulses collected. This is expected behavior for the backends used in this observation (see \citealt{2002ApJ...581..501S} that uses data from PSR~B1534+12 as a demonstration that this kind of integrating-down behavior works efficiently for coherent dedispersion machines in contrast to filterbank machines). {\refr For PSR~J1713+0747, this means that on timescales of $\sim1$\,hr (the largest time which on this timescale we can make multiple samples {\refr with a minimum of eight TOAs}), there is no significant evidence of an absolute noise floor.}

\begin{figure}
  \centerline{\includegraphics[width=1.15\columnwidth]{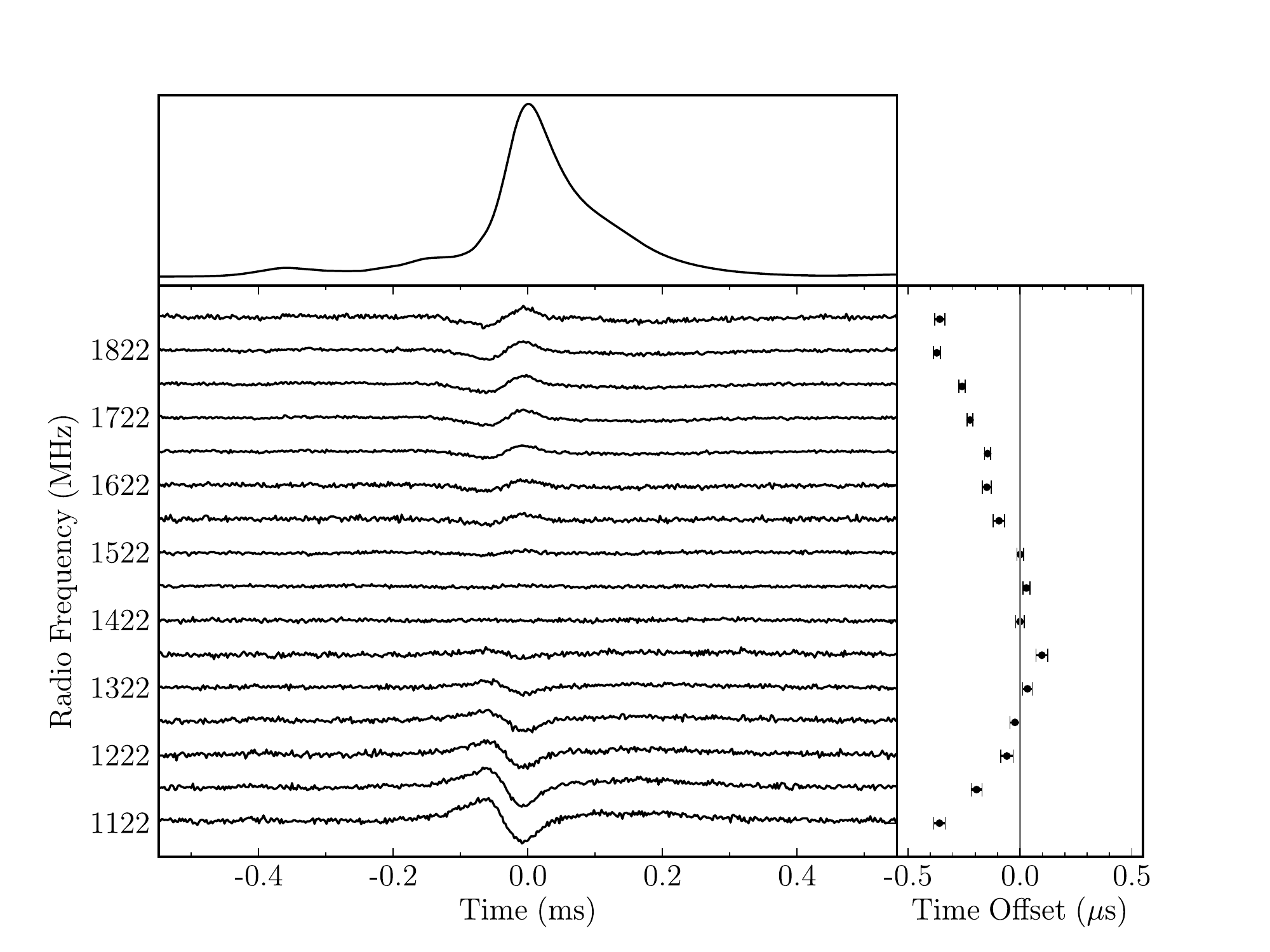}}
 \caption{Differential offset times vs. frequency for the GBT. The top panel shows the standard NANOGrav template used for GUPPI at L-band/1.4\,GHz. The main panel shows the difference profiles as a function of frequency, calculated by subtracting a best-fit template from the data profiles. The right panel shows the mean-subtracted, best-fit phase offsets for each data profile versus frequency. {\refbfb These offsets {\refr are a function of} the FD (frequency dependent) polynomial parameters in \textsc{tempo2} that model timing offsets due to pulse profile frequency evolution. {\refr The shape of the {\refxxxx time offset vs. frequency curve is covariant with any residual dispersion delay across the band.}}}}
 \end{figure}

\begin{figure}
\begin{center}
\centerline{\includegraphics[width=1.15\columnwidth]{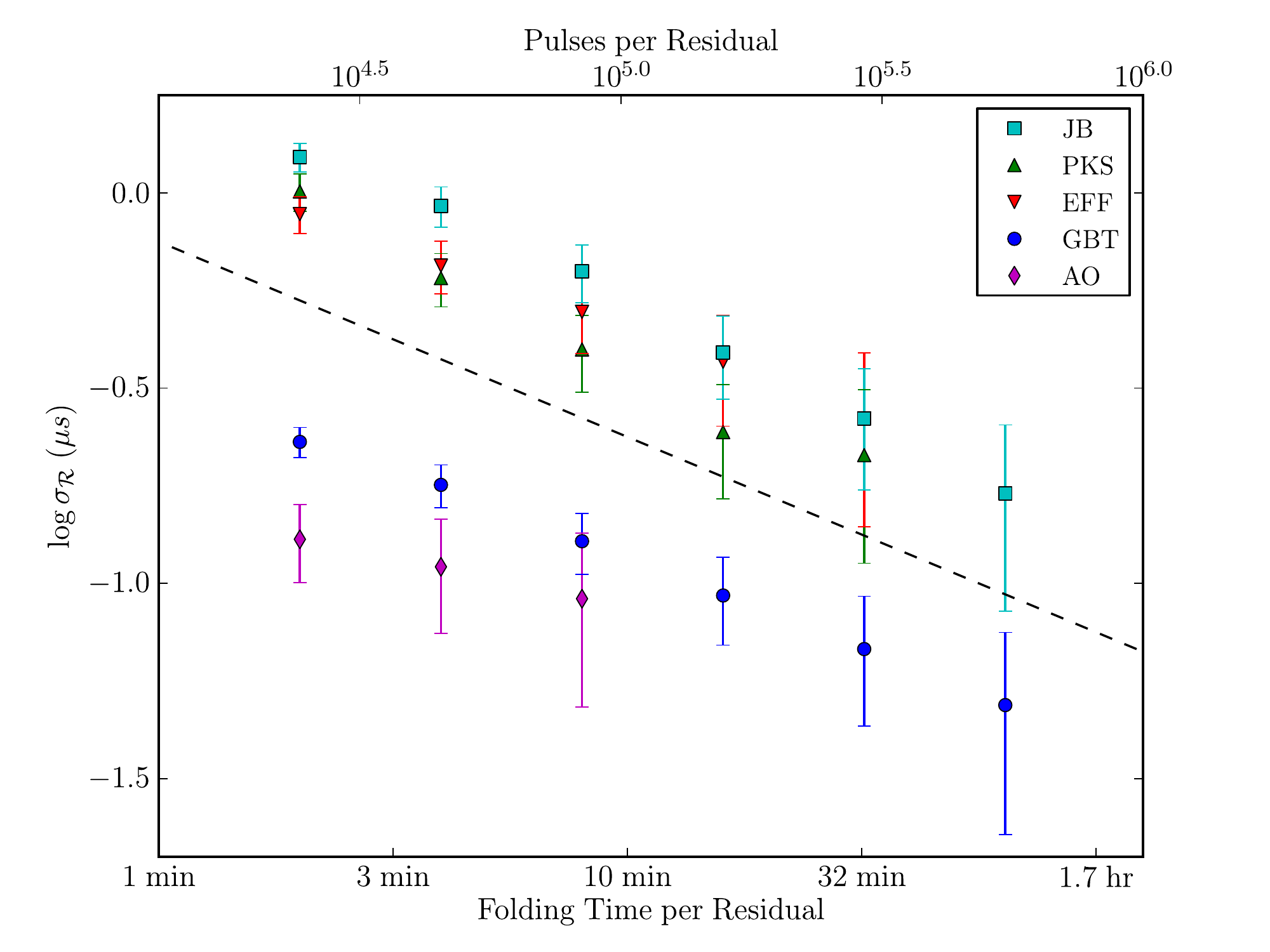}}
\end{center}
\figcaption{Improvement of {\refr L-band/1.4\,GHz} timing rms with number of collected pulses $N$. Both subintegration time per residual and number of pulses are shown in the abscissa for reference. Symbols, from top to bottom: teal squares, Lovell/JB; green upward triangles, Parkes; red downward triangles, Effelsberg; blue circles, GBT; magenta diamonds, Arecibo. {\refr Error bars are the standard error of the scattered rms values.} The dashed lines show a $1/\sqrt{N}$ law for reference. The residuals are derived from telescopes with different bandwidths, and the resulting timing rms values are dependent both on collecting area and bandwidth.}
\end{figure}

The comparative sensitivity of the telescopes can be seen along with the fact that longer tracks produce smaller uncertainties in the residual rms $\sigma_{\mathcal{R}}$. Arecibo and GBT have up to eight times the bandwidth as some of the other telescopes and, {\refbfb considering also the sensitivities,} fall significantly beneath the others in terms of timing rms. Previous studies of PSR~J1713+0747 \citep{2012ApJ...761...64S} have shown a tracking of this $1/\sqrt{N}$ for an $N$ of 1 to 10$^5$, corresponding to subintegration times of {\refbfb 4.57\,ms to 457\,s}.

\subsection{Timing Residual Precision From Template Fitting and Pulse Jitter}
Radiometer noise is always reduced by additional bandwidth but jitter noise, {\refg measurable in high S/N timing observations}, is not improved because it is identical and correlated across frequency. Thus, the minimum expected rms in the broadband timing residuals is somewhat higher than what we would predict from merely reducing the narrowband timing rms by the increased bandwidth factor. The error from jitter, given in \citet{2010arXiv1010.3785C}, is $\sigma_{\textrm{J}} \propto W_{\textrm{eff}}/\sqrt{N}$. The noise value $\sigma_{\textrm{J}}$ has no dependence on bandwidth or telescope sensitivity. Knowing how much the rms {\refr for} an individual TOA would be composed of $\sigma_{\textrm{J}}$ {\refr typically} requires either directly measuring the jitter via single pulses, or measuring the correlated TOAs across frequency. \citet{2012ApJ...761...64S} report a value of 26\,$\mu$s for a single-pulse $\sigma_{\textrm{J}}$ {\refbf from PSR~J1713+0747} also based on Arecibo observations, using both measurement methods. {\refrr \citet{2014MNRAS.443.1463S} have also measured PSR~J1713+0747's single pulse phase jitter rms as $31.1\pm0.7$\,$\mu$s.} Here, we show the presence of jitter in PSR~J1713+0747 by demonstrating the non-dependence of a noise component on telescope sensitivity.}

In order to directly measure the presence of jitter in Figure~6, we binned residuals from \textit{all telescopes} into approximately eight bins/decade in S/N and then took the scatter of the arrival times within each bin, $\sigma_{\mathcal{R}}$. The subintegrations used were 120\,s within frequency bins of $\sim50$\,MHz, {\refr and only residuals for which S/N~$> 1$ were included}. 

{\refg We fit a curve given by the} {\refbfr following equation, which comes from the assumption that the white noise timing residuals, $\sigma_{\mathcal{R}}$, are composed of two other white noise components added in quadrature}:
\begin{equation}
{\refbfb \sigma_{\mathcal{R}} = \sqrt{\sigma_{\textrm{J}}^2 + (\sigma_{S_0}(S_0/S))^2}}
\end{equation}
where $S$ simply represents S/N, $S_0$ refers to a particular fiducial S/N, and $\sigma_{S_0}$ is the timing rms due to template fitting. Using the fiducial 120\,s TOAs, we find that {\refbfb $\sigma_{\textrm{J}} = 0.17\pm0.02\,\mu\mathrm{s}$, which implies a single pulse jitter of $27.0\pm3.3$\,$\mu\mathrm{s}$, consistent with the measurement in \citet{2012ApJ...761...64S} of 26\,$\mu\mathrm{s}$}. {\refrr This value is also consistent within $<2$\,$\sigma$ of the more recent measurement of \citet{2014MNRAS.443.1463S}.} {\refbfb The other fitted parameter, $\sigma_{\textrm{S}_0=1}$, signifying the white noise value in the absence of jitter, was $25.9\pm0.6$\,$\mu\mathrm{s}$.}

\begin{figure*}
\begin{center}
\includegraphics[width=2.2\columnwidth]{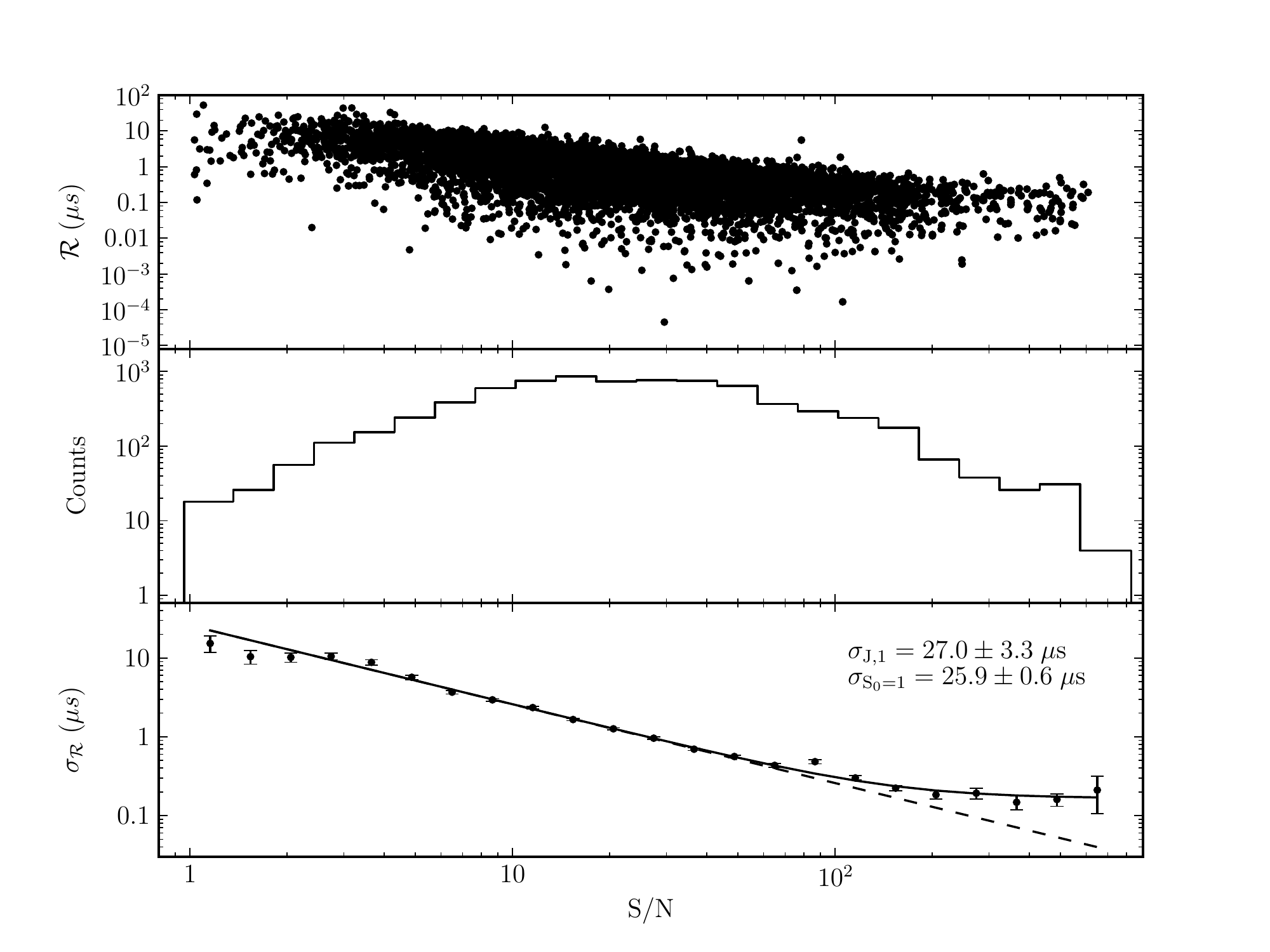}
\end{center}
\figcaption{{\refbf Improvement of {\refbfr residual} rms with S/N, for $\sim50$\,MHz timing residuals. Shown in the top plot {\refxxxx are the timing residual values} as a function of S/N, using data from all telescopes. The middle plot shows the number of residuals in each bin. In the bottom plot, {\refxxxx rms values on the residuals} are shown using logarithmic bins with 8 bins/decade. We fit Equation 3 shown as the solid line fit, which yields a {\refbfr white noise in the timing residuals due to pulse phase jitter of} {\refbfb $\sigma_\textrm{J,1} = 27.0\pm3.3$\,$\mu\mathrm{s}$. Scaled to 120\,s integrations, $\sigma_\textrm{J} = 0.17\pm0.02\,\mu\mathrm{s}$.} All residuals {\refbf shown in the top panel of {\refxx Figure~6}} are for an integration/folding time of 120\,s, {\refr removing residuals below an S/N of 1. The single pulse jitter timing rms is $\sigma_{\textrm{J,1}}$, and $\sigma_{\textrm{S}_0=1}$ is the timing rms in the absence of jitter for a S/N of 1. }}{\refxxxx The dashed line represents the expected timing uncertainties in the absence of pulse phase jitter.}}
\end{figure*}

The presence of jitter in PSR~J1713+0747 does mean that for a telescope as sensitive as Arecibo, LEAP, or GiantLEAP, including the future Square Kilometer Array (SKA) or the Five-hundred-meter Aperture Spherical Telescope (FAST) telescopes, {\refx both $\sigma_{\textrm{J}}$ and the ordinary timing rms scale as $1/\sqrt{N}$} (\citealt{2004NewAR..48.1413C}, \citealt{2010arXiv1010.3785C}, \citealt{2013CQGra..30v4011L}). The dominance of pulse phase jitter seen in the Arecibo portion of the present study may necessitate the use of such long tracks for all future highly-sensitive telescopes to further reduce $\sigma_{\mathcal{R}}$. {\refbfr This is seen in a particularly dramatic fashion over eight of the nine telescopes here.} {\refx Even when using one telescope alone, GBT or Arecibo for instance, the fit to $\sigma_{\mathcal{R}}$ yields $\sigma_\textrm{J,1}$ values of $21.6\pm4.1\,\mu\mathrm{s}$ (GBT) and $27.9\pm5.3\,\mu\mathrm{s}$ (Arecibo). The $\sim1$\,$\sigma$ consistency of each single-telescope value with the all-telescope value implies that the jitter numbers reported are not telescope-dependent, and are intrinsic to the pulsar as expected.}

Single pulse phase jitter causes a timing error $\propto 1/\sqrt{N}$ that is independent of S/N. {\refg The two contributions are equal for single-pulse (S/N)$_1 \sim1$ \citep{2012ApJ...761...64S}.} S/N~$> 1$ single pulses {\refbfr should be present, given {\refxx that a telescope is sensitive}} to jitter noise \citep{2010arXiv1010.3785C}. 

A more detailed analysis is deferred to a separate publication. Single pulses can be extracted from any of the observations with baseband data or intensity integrations, which {\refbfr were taken at various times} during the global campaign in all nine telescopes. However, it is interesting that fitting Equation 3 across the eight telescopes in {\refxx Figure~6} allows a jitter measurement without probing into residuals with a subintegration time of {\refxxxx $< 120$\,s}, much less with single pulses. {\refr The residuals in the brightest bin, rescaled to subintegrations of a single period, correspond to a single pulse S/N of $\sim3$.}

\subsection{Strong Correlation of Diffractive Scintillation Between Telescopes}

Figure~3 shows that the frequency-time structure in the dynamic
spectrum is qualitatively identical between simultaneous measurements from
different telescopes, apart from low-elevation-angle observations and from {\refbfr masked} episodes of RFI.   This high correlation includes  telescope pairs with the largest separations (up to 9000 km), Parkes and GMRT; GMRT with the European telescopes (Jodrell Bank, Effelsberg, and WSRT), between the GBT and the European telescopes, and between the GBT and Arecibo.

The observations are consistent with the expectation that the dynamic spectra
for PSR~J1713+0747 should be highly correlated between all terrestrial
telescopes because of the low level of scattering along the line of sight.
We estimate the spatial scale $\ell_d$ of the
diffraction pattern from the parallax distance $d$ = $1.05\pm0.06$\,kpc
\citep{2009ApJ...698..250C} and the scintillation bandwidth
$\Delta\nu_d \approx 0.6\pm0.2$\,MHz at 0.43\,GHz
\citep{2002ApJ...581..495B}
using Eq.\,9 of \citet{1998ApJ...507..846C},
\begin{eqnarray}
\ell_d = \frac{1}{\nu}\left(\frac{cd\Delta\nu_d}{4\pi C_1} \right)^{1/2},
\end{eqnarray}
where $C_1 = 1.16$ using a default, uniform Kolmogorov scattering medium.
This yields $\ell_d \approx 5\times 10^4$\,{\rm km} at 0.43\,GHz and scaling
by $\nu^{1.2}$, the diffraction scale at 1.4\,GHz is
$\ell_d \approx 2\times10^4$\,km, much larger than the Earth.

\section{Initial Results on the Noise Budget of the Timing Residuals}
In this section we briefly consider some aspects of {\refc PSR~J1713+0747's} noise budget, in other words, whether the S/N across telescopes corresponds to expectations from general considerations. A more detailed consideration of the noise budget will be found in a forthcoming paper.

Figure~7 (top panel) shows the grand average profile for {\refxxxx the} eight telescopes that observed at L-band/1.4\,GHz. Low frequencies are shown in the lower panels. Profiles are summed across the full band from individual subbands' residual values, weighted by the off-pulse noise values, {\refxxxx and folded according to the measured L-band/1.4\,GHz value of DM}. The resulting L-band/1.4\,GHz profile has a S/N of $\sim4000$, where the signal value is taken as the amplitude at the maximum value of the summed pulse and the noise is taken from the off-pulse part of the combined profile. S/N values were calculated using the first 100 bins {\refbfr of the profile} for the noise region. {\refg The profile was centered at maximum, with $N_{\phi}$ = 512, before summation. {\refr See Table~1 for estimates of the degree to which each telescope contributes to the total pulse profile S/N.}} {\refrrrr Low frequency profiles are also shown for reference, showing significant profile evolution with frequency.} {\refrrrrr The pulsar is weak at low frequencies because it
appears to turn over somewhere above the LOFAR band (\citealt{Has14} in preparation).}

The minimum rms on timing residuals from the eight-telescope L-band/1.4\,GHz profile can be estimated using Equation\,\ref{eq:eq1}. Given an effective {\refbfg pulse} width of 0.54\,ms, that we have 512 phase bins, and that the {\refx 24-hr} pulse profile S/N in the grand average profile was about 4000, {\refbfg this yields a template fitting error, $\sigma_{\mathrm{S/N}}$, of about 3\,ns. Given a {\refx 24-hr} implied template fitting error of 3\,ns and an implied {\refx 24-hr} jitter timing error of 170/$\sqrt{30}/\sqrt{24}$ = 6.3\,ns (rescaling from the 120\,s jitter value given in Section~3.2 to a 24\,hr value), we add these values in quadrature to arrive at an approximate timing uncertainty of 7\,ns. Jitter and template fitting alone would then yield a timing residual error of 7\,ns on a {\refx 24-hr} TOA.

The uncertainty from our L-band/1.4\,GHz DM measurement of 0.00001\,pc\,cm$^{-3}$ from Section~2 corresponds to 24\,ns of smearing across the band. A better measurement of the DM on MJD 56465 -- 56466 would require incorporation of the 150\,MHz LOFAR data and the 350\,MHz WSRT data, but taking into account a model of the pulse profile evolution with frequency that extends to these lowest two frequencies. Such a model is a topic of exploration for a future paper on interstellar electron density variations. Any discovered DM variations, along with an improved DM smearing value, would need to inform the noise floor assessment.}

\begin{figure}
  \begin{center}
  \centerline{\includegraphics[width=1.15\columnwidth]{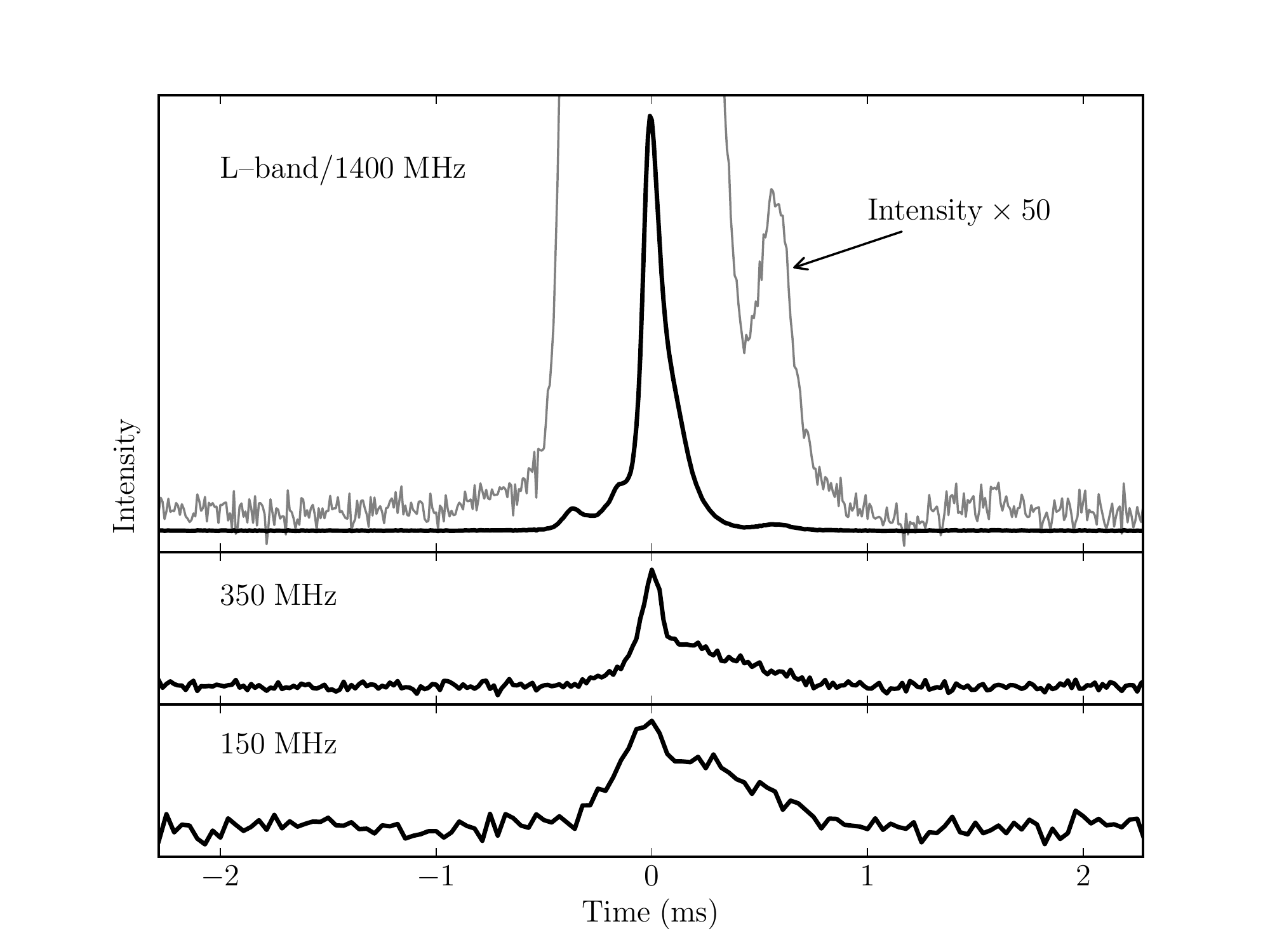}}
  \end{center}
 \figcaption{The grand average profile for all telescopes. The top panel was created across all bandwidths from those telescopes which observed at L-band/1.4\,GHz. While it has been shown in Section~2 that there is some profile smearing occurring due to the pulse profile evolution with frequency, we sum in weighted fashion from individual subbands' residual values. The resulting profile has a S/N of $\sim4000$, where the signal value is taken as the amplitude at the maximum value of the summed pulse and the noise is taken in the first 100 phase bins of the off-pulse part of the combined profile. {\refbfr The bottom two panels show the grand average profiles for low frequencies, manifesting the significant profile evolution with frequency.} {\refxxxx The DM for all telescopes is set by the fitted L-band/1.4\,GHz DM.}}
 \end{figure}
\def\be{\begin{equation}}
\def\ee{\end{equation}}
\def\ba{\begin{eqnarray}}
\def\ea{\end{eqnarray}}

\newcommand{\Weff}{\ensuremath{W\mathrm{_{eff}}}\xspace}
\newcommand{\Dt}{\ensuremath{{\Delta t}}\xspace}
\newcommand{\Dto}{\ensuremath{{\Delta t_0}}\xspace}
\newcommand{\DtJ}{\ensuremath{{\Delta t_J}}\xspace}
\newcommand{\niss}{\ensuremath{{n\mathrm{_{ISS}}}}\xspace}
\newcommand{\SNR}{\mathrm{SNR}\xspace}
\newcommand{\half}{\frac{1}{2}\xspace}

\newcommand{\nocomma}{}
\newcommand{\smathbf}[1]{\ensuremath{\boldsymbol{#1}}}

\section{Further Work}

The initial results presented will be important {\refbfr for the three PTAs and for the IPTA as a whole. For some telescopes, PSR~J1713+0747 is timed (or is under consideration to be timed) for longer observation durations at each epoch, or at a higher observing cadence. Being amongst a small set of pulsars with the lowest timing residual rms values, it strongly influences the sensitivity of the entire IPTA, despite the necessity of calculating angular correlations in order to populate the Hellings and Downs diagram and detect a} stochastic background of GWs \citep{1983ApJ...265L..39H}. Increasing the observation duration for these pulsars helps the sensitivity of the IPTA to other types of GW source populations, such as burst, continuous wave, and memory bursts (see \citealt{2014ApJ...794..141A} for current limits on continuous wave sources). {\refbfr The first results presented in this work provide a starting point on the subtleties that may emerge with an increasing dependence on {\refc PSR~J1713+0747} and similar pulsars such as PSR~J0437$-$4715 and {\refc PSR~J1909$-$3744.}}

We plan to release papers on the following subjects, among others, based on this dataset:

\textit{The Noise Budget of the 24-Hour Global Observation of PSR~J1713+0747}: the question to be explored here is the degree to which one can dissect the noise present on the different timescales relevant in this observation. From single pulses at the $\mu$s resolution all the way to the full 24\,hr, {\refbfr the statistical structure of noise in timing residuals} can be probed using various diagnostics. Structure on different timescales can be probed by looking at the pulsar with the Allan variance function. Single pulses can also be exploited in order to search for smaller timescale structure such as giant pulses, mode changes, and drifting sub-pulses (see \citealt{2010ApJ...725.1607S}). Shape changes can be probed and possibly mitigated using various methods (\citealt{1993ASPC...36...43C}, \citealt{2007PhDT........14D}, \citealt{2011MNRAS.418.1258O}). 

\textit{Interstellar Electron Density Variations {\refc and Pulse Profile Frequency Evolution}}: the all-telescope dynamic spectrum can yield interesting information in a further analysis. Given the data obtained at low-frequencies with LOFAR and WSRT, it will be informative to search for correlations between events occurring in the L-band/1.4\,GHz dynamic spectrum and the highly scattered structure at 150\,MHz and 350\,MHz respectively. {\refc Analysis can be done using the LOFAR and WSRT data to obtain more accurate DM measurements, while taking into account the significant profile evolution between the two low-frequency observations and the L-band/1.4\,GHz observations.}

\textit{GiantLEAP}: one of the signature objectives of this observation is to use the {\refrr European} telescopes, Arecibo, the GMRT, and the GBT as a single phased array, or at least to expand LEAP with some subset thereof. In particular, RFI excision using simultaneous data from a subset of telescopes might significantly improve the quality of the phased array over one more locally situated. Once the proper correlations are performed, in principle the timing rms of PSR~J1713+0747 from the largest simultaneous collecting area ever used will be obtained. However, what practical limitations will come into play at realizing this ideal would be the subject of future studies. Undoubtedly, whatever timing results will be obtained will be highly affected by the presence of pulse phase jitter.

\textit{Polarization Studies.} Most telescopes in this study took {\refbfr polarimetric} data (see Appendix) and studying the timescales of PSR~J1713+0747's polarization over the 24 hours could provide new insights, particularly at the single pulse level.

\section{Conclusions}

We have presented an overview of the goals and data products of the 24-hour global campaign on PSR~J1713+0747. This $\sim$\,60\,TB dataset is useful for many goals which will be explored in further papers, including but not limited to: better determination of the overall noise budget for PTAs, a wide-bandwidth, long-timespan examination of the effects of the ISM on pulsar timing, combining baseband data from simultaneous observations for the GiantLEAP experiment, an examination of single pulses and their phenomenology over the 24 hours, and many others. 

In the first results presented here, some interesting conclusions can already be drawn. PSR~J1713+0747's intrinsic pulse phase jitter ($\sim27.0$\,$\mu$s for single pulses) can be measured by fitting a noise model across all telescopes, even when {\refrrrrr TOA integration times are as long} as 120\,s. {\refq The improvement of timing residual rms is not found to depart significantly from a factor of $1/\sqrt{N}$, where $N$ is the number of integrated pulses.} Finally, the diffraction scale at 1.4\,GHz was seen to be $\ell_d \approx 2\times10^4$\,km, much larger than the Earth, from the overlapping scintillation pattern seen in the dynamic spectrum in Figure~3.

\section{Acknowledgments}

{\refxx We thank the IPTA for its ongoing international collaboration that motivated this project, which would have been challenging for any PTA consortium to execute alone.} We would {\refxx especially} like to thank {\refxxxx NSF-PIRE, the EPTA, and the PPTA} for enabling the IPTA 2012 Meeting in Kiama, New South Wales, Australia, and the IPTA 2013 Meeting in Krabi, Thailand. During both these meetings, the unique synergy of many IPTA personnel interacting in the same place brought about both the idea for, and later, the execution of, this observation. We are grateful to the University of Sydney and the National Astronomical Research Institute of Thailand (NARIT) for hosting these meetings.

The work of SJC, JMC, PBD, TD, FJ, GJ, MTL, TJWL, JL, DRM, MAM, NP, SMR and KS was partially supported through the National Science Foundation (NSF) PIRE program award number 0968296. PL acknowledges the support of IMPRS Bonn/Cologne and FQRNT B2. We thank the telescope schedulers Hector Hernandez, Alex Kraus, Tony Minter, and many others for working hard to ensure that this observation was given adequate time, given the difficulty of scheduling nine telescopes for simultaneous observations. We thank the staff of the GMRT who have made these observations possible. The GMRT is run by the National Centre for Radio Astrophysics of the Tata Institute of Fundamental Research. {\refrrrr NANOGrav research at University of British Columbia is supported by an NSERC Discovery Grant and Discovery Accelerator Supplement and by the Canadian Institute for Advanced Research.} The National Radio Astronomy Observatory is a facility of the NSF operated under cooperative agreement by Associated Universities, Inc. The Arecibo Observatory is operated by SRI International under a cooperative agreement with the NSF (AST-1100968), and in alliance with Ana G. M\'{e}ndez-Universidad Metropolitana, and the Universities Space Research Association. The 100-m Effelsberg telescope is operated by the Max-Planck-Institut f\"ur Radioastronomie (MPIfR). LOFAR, the
Low Frequency Array designed and constructed by ASTRON, has
facilities in several countries, that are owned by various parties
(each with their own funding sources), and that are collectively
operated by the International LOFAR Telescope (ILT) foundation under a joint scientific policy. Access to the Lovell Telescope is supported through an STFC rolling grant. The Nan\c cay radio telescope is part of the Paris Observatory, associated with the Centre National de la Recherche Scientifique (CNRS), and partially supported by the R\'{e}gion Centre in France. The Parkes Radio telescope is part of the Australia Telescope National Facility which is funded by the Commonwealth of Australia for operation as a National Facility managed by CSIRO. The Westerbork Synthesis Radio Telescope is operated by the Netherlands Foundation for Research in Astronomy (ASTRON) with support from the NWO (Netherlands Organisation for Scientific Research). Some of this work was supported by the ERC Advanced Grant ``LEAP'', Grant Agreement Number 227947 (PI M. Kramer). Part of this research was carried out at the Jet Propulsion Laboratory, California Institute of Technology, under a contract with the National Aeronautics and Space Administration. Analysis of LOFAR
data was supported by ERC Starting Grant ``DRAGNET'' (337062; PI
J.\,Hessels).

{\refxx Finally, we thank the anonymous referee for the helpful suggestions.}
\begin{center}
\begin{deluxetable*}{lccc}[c]
\tablecolumns{4}
\tablecaption{Timing model parameters}
\tablehead{
\colhead{Parameter} & \colhead{Value} & \colhead{Held Fixed in Fit?$^\textrm{a}$} & \colhead{Parameter Uncertainty}
}
\startdata
Right Ascension, $\alpha$ (J2000)	&	17:13:49.5331497	&	Y	&	$5\times10^{-7}$	\\
Declination, $\delta$ (J2000)	&	07:47:37.492844	&	Y	&	$1.4\times10^{-5}$	\\
Proper motion in $\alpha$, $\nu_{\alpha}$ (mas yr$^{-1}$)  	&	4.922	&	Y	&	0.002	\\
Proper motion in  $\delta$, $\nu_{\delta}$ (mas yr$^{-1}$)	&	$-3.909$	&	Y	&	0.004	\\
Parallax, $\pi$ (mas) 	&	0.88	&	Y	&	0.03	\\
Spin Frequency (Hz)	&	218.81184381090227 	&	Y	&	$7\times10^{-14}$	\\
Spin down rate (Hz$^{2}$)	&	$-4.083907\times10^{-16}$	&	Y	&	8$\times10^{-22}$	\\
Reference epoch (MJD)	&	54971	&	Y	&		\\
Dispersion Measure (pc cm$^{-3}$)	&	15.99113	&	N	&	$1\times10^{-5}$	\\
Profile frequency dependency parameter, FD1	&	$1.328\times10^{-5}$	&	Y	&	$4\times10^{-8}$	\\
Profile frequency dependency parameter, FD2	&	$-3.73\times10^{-5}$	&	Y	&	$2\times10^{-7}$	\\
Profile frequency dependency parameter, FD3	&	$3.24\times10^{-5}$	&	Y	&	$7\times10^{-7}$	\\
Profile frequency dependency parameter, FD4	&	$-1.07\times10^{-5}$	&	Y	&	$5\times10^{-7}$	\\
Solar System ephemeris	&	DE421	&	Y	&		\\
Reference clock	&	TT(BIPM)	&	Y	&		\\
Binary Type	&	T2$^\textrm{b}$	&	Y	&		\\
Projected semi-major axis, x (lt-s)	&	32.34242245	&	Y	&	$1.2\times10^{-7}$	\\
Eccentricity, e	&	$7.49414\times10^{-5}$	&	Y	&	$6\times10^{-10}$	\\
Time of periastron passage, T$_0$ (MJD)	&	54914.0602	&	Y	&	0.0003	\\
Orbital Period, P$_b$ (day)	&	67.825147	&	Y	&	$5\times10^{-6}$	\\
Angle of periastron, $\omega$ (deg)	&	176.1978	&	Y	&	0.0015	\\
Derivative of periastron angle, $\dot{\omega}$ (deg)	&	0.00049	&	Y	&	0.00014	\\
Companion Mass, M$_c$ (M$_{\sun}$)	&	0.29	&	Y	&	0.01	
\footnotetext{\refg We also fit for arbitrary jumps between telescopes, which are not astrophysical and not shown here.}
\footnotetext{\citet{1986AIHS...44..263D}}
\end{deluxetable*}
\end{center}
\appendix
\section{Additional Observing Details}

Most information about the observation can be found in Table~1. {\refbfr \textsc{Tempo2} parameters are in this Appendix in Table\,3, most of which derive from the parameters calculated in \citealt{Zhu15} (in preparation).}
\section{Effelsberg 100-m Radio Telescope}
Effelsberg's data taking began with 9.3\,hr of contiguous observing at 1380\,MHz. The PSRIX instrument (\citealt{Kar14}, in preparation) was used for both baseband and folding modes (after real-time coherent dedispersion). There were two baseband sessions, one 30\,min and the other 1\,hr. In folding mode, PSRIX was configured to coherently dedisperse and fold $8\times25$\,MHz bands. Each resulting file has 200\,MHz of bandwidth (though some channels are {\refbfr removed due to RFI}), 128 channels, 1024 phase bins of 4.47\,$\mu$s each, 10\,s subintegrations, and full polarization information. In total there were $\sim6$\,hr (2.1\,GB) of folding mode data. In baseband mode, data was recorded as $8\times16$\,MHz (128\,MHz) subbands in order to be compatible with other LEAP telescopes. {\refbfg The data were flux calibrated using the noise diode, which in turn was calibrated using a North-On-South triplet of observations of 3C\,218 following the {\refx 24-hr} campaign. The data were recorded in ``Timer Archive'' format. They were converted to \textsc{psrfits} format as part of Effelsberg's standard data reduction pipeline.}
\section{Giant Meterwave Radio Telescope}
The GMRT {\refbfg used 22 antennas, employing} two observing modes -- a total {\refrrrr offline} coherent filterbank mode with 65.1\,kHz spectral and 61.44\,$\mu$s time resolution, and a coherent array voltage mode with a single subband for the baseband portion of the observation. {\refbfr Both these modes are described in \citet{2010ExA....28...25R}.} The frequency range was from 1371\,MHz to 1404\,MHz. There were $7\times1$\,hr recording scans interleaving with phasing scans for the array, as well as a 50\,min coherent array baseband voltage recording scan. This resulted in 436\,GB of raw filterbank data and 460\,GB of voltage data. {\refbfr The GMRT filterbank data (61.44\,$\mu$s time resolution) are 16-bit and in a format compatible with the \textsc{presto}\footnote[2]{http://www.cv.nrao.edu/${}_{\textrm{\symbol{126}}}$sransom/presto/} searching suite. The GMRT coherent array voltage data (15\,ns time resolution) are 8-bit and in a DSPSR friendly format.} {\refbfr The GMRT coherent array provides some built-in immunity to RFI as the processing pipeline adjusts the antenna phases to correct for the effect of rotation of the sky signals, which in turn de-correlates the terrestrial signals. Interleaved calibrator observations ({\refr QSO\,J1822--096} in this case) every 2\,hr were required to optimize the coherent array sensitivity at the observing frequencies. The antenna-based gain offsets are also corrected using this calibrator before making the coherent beam.}
\section{Lovell Telescope at Jodrell Bank Observatory}
{\refc L-band/1.4\,GHz observations of PSR~J1713+0747 were obtained with the Lovell
telescope at Jodrell Bank over an 11.5\,hr timespan. The data are
continuous except for a few brief gaps due to the telescope being
parked for wind constraints. Two instruments were used; i) the DFB
performed real-time folding with incoherent dedispersion, producing
folded 10\,s subintegrations of 1024 pulse phase bins of 4.47\,$\mu$s
in size, 0.5\,MHz channels over a 384\,MHz wide band centered at
1532\,MHz, ii) the ROACH, using a Reconfigurable Open Architecture
Computing Hardware FPGA board performing real-time folding with
coherent dedispersion using the \textsc{psrdada}\footnote[3]{http://psrdada.sourceforge.net} and \textsc{dspsr}
software packages. This provided 10\,s
subintegrations with 2048 pulse phase bins of 2.23\,$\mu$s in size and
covered 400\,MHz wide band centered at 1532\,MHz, split into 25
subbands of 16\,MHz, each channelized to provide 0.25\,MHz
channels. The ROACH was also used to record baseband data during the
times when Nan\c cay and Arecibo observed the pulsar. Dual
polarization, Nyquist sampled baseband data, 8-bits digitized, was
recorded for the lower 8 subbands of 16\,MHz (1332 to 1460\,MHz),
while the remaining seventeen 16\,MHz subbands performed real-time
folding with coherent dedispersion as before. At the end of the
observations the baseband data was folded and coherently dedispersed
with the same parameters as for the real-time folding to give one
continuous observing run. The spectral kurtosis method by Nita et
al.\,(2007) for identifying and flagging RFI, as implemented
in \textsc{dspsr}, was used to excise RFI in real time. After the
observations a combination of manual and automatic RFI excision was
performed to clean the data further. The folded profiles were
polarization calibrated using the Single Axis model \citep{2004ApJS..152..129V} using observations
of the noise diode and observations of pulsars with known polarization
properties.}
\section{Low Frequency Array}
LOFAR observed from 110 -- 190\,MHz using the BG/P beam-former and
correlator (see \citealt{2013A&A...556A...2V}). The sampling time was 5.12\,$\mu$s and 400 subbands of 0.195\,MHz each were recorded. Full polarization
information was taken in complex voltage mode; see \citet{2011A&A...530A..80S} for more information on pulsar observing modes with
LOFAR. {\refxxx The raw data volume (32-bit) was
4.5\,TB/hr, yielding 40\,TB of complex-voltage raw data (of which only
1\,hr, 4.5\,TB, of raw data has been archived long-term; the rest is
only available as folded archives, as summarized in Table\,1)
The 40\,TB value not included in total the 60\,TB value for
the all-telescope data.} These data were coherently dedispersed and
folded offline using {\refxxx \textsc{dspsr}}.  The following 23 LOFAR Core stations were
combined for the {\refxxx 9-hr} observation: CS001, CS002, CS003, CS004,
CS005, CS006, CS007, CS011, CS017, CS021, CS024, CS026, CS028, CS030,
CS031, CS032, CS101, CS103, CS201, CS301, CS302, CS401, and CS501.
See \citet{2013A&A...556A...2V} for more specific location information;
by default the phase center {\refxxx of the tied-array beam} is placed at the position of CS002.
\section{Nan\c cay Decimetric Telescope}
{\refbfg Nan\c cay observed using the NUPPI (Nan\c cay Ultimate Pulsar Processing Instrument) backend at L-band/1.4\,GHz with a total bandwidth of 512 MHz, split into {\refxxxx $32\times16$}\,MHz channels and 8-bits digitized. The profiles were folded and integrated over 1\,min and finally stored in a 29\,MB \textsc{psrfits} file. All data were coherently dedispersed and the total Nan\c cay observation lasted for $\sim1$\,hr. The \textsc{psrchive} program \textsc{pac} was used to do the polarization calibration with the Single Axis model and automatic zapping was then applied with the \textsc{paz} program. The TOAs were produced with \textsc{pat} using a high S/N template.}
\section{Parkes 64\,m Telescope}
The Parkes 64\,m radio telescope observed PSR~J1713+0747 at 1362\,MHz using both fold-mode and baseband mode in parallel for  $\sim6$\,hr. This time included four 3\,min noise diode calibration scans, between 64\,min blocks of folding. The backends DFB3/4 (incoherent filterbank, 60\,s foldings, 1024 frequency channels over 256\,MHz bandwidth, 1024 phase bins of 4.47\,$\mu$s each), APSR ({\refxx real-time} coherent filterbank, 30\,s foldings, 512 frequency channels over 256\,MHz bandwidth, 1024 phase bins of 4.47\,$\mu$s each), and CASPSR (30\,s foldings, 400\,MHz bandwidth with $\sim10$\,MHz band edges, 1024 phase bins of 4.47\,$\mu$s each) observed in parallel, allowing for simultaneous baseband and folding mode observations. {\refbfr RFI was removed, and consistent results were obtained with all backends using a median filter in the frequency domain. Polarization and flux calibration for the DFB data used} the standard monthly flux calibrations on the Hydra A radio galaxy \citep{2013PASA...30...17M}. CASPSR was calibrated for differential gain and phase but nothing else. 
{\refbfr RFI was mitigated using the
radio-frequency domain filter implemented in the \textsc{psrchive} command \textsc{paz}
\citep{2004PASA...21..302H}.  Additionally, the CASPSR
instrument mitigates RFI in real time by rejecting portions of the
data that show distribution inconsistent with receiver noise by using
a spectral kurtosis filter.
Digital filterbank data were calibrated for cross coupling using a
model for the feed derived from long-track observations of the bright
source PSR~J0437$-$4715.
Observations over these wide parallactic angles enable the measurement
of the feed cross coupling and ellipticity to be measured \citep{2004ApJS..152..129V}.
The model used was the average of many long-track observations because
the feed parameters were not found to change significantly with time.
CASPSR observations were not calibrated for this cross coupling.
PSR~J1713+0747 is only a modestly polsarized pulsar and the effects of
correction for these effects were found to be negligible improvement
on its long term timing precision \citep{2013PASA...30...17M}.
Details of this calibration are further described in \citet{2013PASA...30...17M}.}
All the data are digitised with 8-bit digitisers. The baseband dataset is 30\,TB, with 7.5\,GB for the fold-mode dataset.
The fold mode used the \textsc{psrfits} data format, and the baseband mode used the \textsc{psrdada} format.
\section{Westerbork Synthesis Radio Telescope}
{\refc WSRT is a $14\times25$\,m dish East-West array, which for pulsar observations is used in tied-array mode and is phased up before the observations for all observing bands used. There were no absolute flux calibrations done for these observations. Full polarization information was stored. Due to the array-nature of the telescope there is usually very little RFI and therefore any remaining leftover narrow-channel RFI zapping is done offline using \textsc{psrchive} tools. For these observations we had 11 out of the total 14 dishes available.}
The WSRT observations made use of PuMa-II instrument \citep{2008PASP..120..191K}, observing at both L-band/1.4\,GHz and 350\,MHz each with 8 separate bands, which are each either 10\,MHz wide (for the 350\,MHz observations) or 20\,MHz wide (for the 1380\,MHz observations). 
{\refrr The individual bands overlapped by 2.5\,MHz of the
10\,MHz bandwidth for the 350\,MHz observations. At L-band/1.4\,GHz the bands
overlapped by 4\,MHz out of the 20\,MHz bandwidth to ensure full
overlap with the 16\,MHz bands at the Effelsberg, Lovell and Nan\c cay telescopes and allow for coherent addition of the data.}
For each there are 10\,s subintegrations stored with 256 bins of 17.9\,$\mu$s each across the profile and 64 channels across each band (which is 10 or 20\,MHz depending on the frequency). The low frequency TOAs were made using \textsc{psrchive} templates, created using the \textsc{paas} (analytic template) routine based on a high-S/N summation of many other observations. 
\section{Arecibo Observatory and the NRAO Green Bank Telescope}
Both Arecibo and the GBT, as mentioned earlier, used non-folded intensity recording for their non-baseband portions of the observation so that single-pulse data would be available over the entire span of the observation, without the cost in bandwidth. Unlike baseband data, no voltage, and thus no electromagnetic phase information, is present. This observation mode is essentially a pulsar search mode with coherent dedispersion, according to the source's DM. Calibrations at the start of the observations at both GBT and Arecibo were performed with a noise diode switched at 25\,Hz, including the polarization calibration. Absolute flux measurements, also including the polarization calibration, were performed on {\refr QSO}\,B1442+101 at the GBT and {\refr QSO}\,J1413+1509 at Arecibo. {\refbf We apply these calibrations via the Single Axis model using the the \textsc{psrchive} program \textsc{pac}.}
\subsection{Green Bank Telescope}
GBT began its observation starting with baseband mode, switching to an hour of intensity integration observing mode, switching back to baseband mode for 30\,min, and then returning to intensity integration mode for the remaining seven hours. The two baseband sessions would be simultaneous with other telescopes. The intensity integration mode had {\refxx $256\times3.125$\,MHz} frequency channels and had a time resolution of 5.12\,$\mu$s. The same was planned for the PUPPI backend at Arecibo. Full polarization information was recorded. Intensity integrations are effectively search mode but coherently dedispersed with the known DM. During the switch from the first baseband time block to the first intensity integration time block, some time was lost due to a problem with the observing mode on GUPPI, and so to ensure a safer data rate, the observing mode was switched to {\refxx $128\times6.25$\,MHz} channels with a time resolution of 2.56\,$\mu$s. For this reason, the settings on GUPPI were different from PUPPI. The tradeoff is that with slightly wider channels in PUPPI, although yielding a better set of phase bins, cannot in principle excise RFI as efficiently. This interruption and restart caused a small gap in the data, visible in Figure~3. {\refq GBT's backend employing real-time cyclic spectroscopy (\citealt{Jon15}, in preparation) was also used in parallel during the folding mode observations, with 65\,MHz of bandwidth centered at 1398\,MHz.}
\subsection{Arecibo Observatory}
Finally, Arecibo joined the observation. Beginning with 30\,min of baseband observing in order to contribute to GiantLEAP, it them switched over to PUPPI intensity integration mode, a different mode used than that at GBT as just described.
\shortauthors{Dolch \& Lam et al.}

\bibliographystyle{apj}
\bibliography{apjmnemonic,1713global}

\begin{thebibliography}{}

\bibitem[Archibald et al.(2014)]{2014ApJ...790L..22A} Archibald, A.~M., 
Kondratiev, V.~I., Hessels, J.~W.~T., 
\& Stinebring, D.~R.\ 2014, \apjl, 790, L22 

\bibitem[\protect\citeauthoryear{{Arzoumanian} et~al.}{{Arzoumanian}
  et~al.}{2014}]{2014ApJ...794..141A}
{Arzoumanian}, Z., et~al. 2014, \apj, 794, 141 

\bibitem[\protect\citeauthoryear{{Arzoumanian} et~al.}{{Arzoumanian} et~al.}{2015}]{Arz15}
{Arzoumanian}, Z., et~al. 2015, {\it in preparation}

\bibitem[\protect\citeauthoryear{{Bassa} et~al.}{{Bassa} et~al.}{2014}]{Bas14}
{Bassa}, C., et~al. 2014, {\it in preparation}

\bibitem[\protect\citeauthoryear{{Bogdanov} et~al.}{{Bogdanov}
  et~al.}{2002}]{2002ApJ...581..495B}
{Bogdanov}, S., {Pruszy{\'n}ska}, M., {Lewandowski}, W.,  \& {Wolszczan}, A.
  2002, \apj, 581, 495

\bibitem[\protect\citeauthoryear{Caprini, Durrer, \& Siemens}{Caprini
  et~al.}{2010}]{PhysRevD.82.063511}
Caprini, C., Durrer, R.,  \& Siemens, X. 2010, Phys. Rev. D, 82, 063511

\bibitem[\protect\citeauthoryear{{Chatterjee} et~al.}{{Chatterjee}
  et~al.}{2009}]{2009ApJ...698..250C}
{Chatterjee}, S., et~al. 2009, \apj, 698, 250

\bibitem[Cognard 
\& Backer(2004)]{2004ApJ...612L.125C} Cognard, I., \& Backer, D.~C.\ 2004, \apjl, 612, L125 

\bibitem[\protect\citeauthoryear{{Cordes}}{{Cordes}}{1993}]{1993ASPC...36...43C}
{Cordes}, J.~M. 1993, in Astronomical Society of the Pacific Conference Series,
  Vol.~36, Planets Around Pulsars, ed. J.~A. {Phillips}, S.~E. {Thorsett}, \&
  S.~R. {Kulkarni}, 43

\bibitem[Cordes(2013)]{2013CQGra..30v4002C} Cordes, J.~M.\ 2013, Classical 
and Quantum Gravity, 30, 224002 

\bibitem[\protect\citeauthoryear{{Cordes} \& {Downs}}{{Cordes} \&
  {Downs}}{1985}]{1985ApJS...59..343C}
{Cordes}, J.~M.,  \& {Downs}, G.~S. 1985, \apjs, 59, 343

\bibitem[\protect\citeauthoryear{{Cordes} \& {Jenet}}{{Cordes} \&
  {Jenet}}{2012}]{2012ApJ...752...54C}
{Cordes}, J.~M.,  \& {Jenet}, F.~A. 2012, \apj, 752, 54

\bibitem[\protect\citeauthoryear{{Cordes} et~al.}{{Cordes}
  et~al.}{2004}]{2004NewAR..48.1413C}
{Cordes}, J.~M., {Kramer}, M., {Lazio}, T.~J.~W., {Stappers}, B.~W., {Backer},
  D.~C.,  \& {Johnston}, S. 2004, \nar, 48, 1413

\bibitem[\protect\citeauthoryear{{Cordes} \& {Rickett}}{{Cordes} \&
  {Rickett}}{1998}]{1998ApJ...507..846C}
{Cordes}, J.~M.,  \& {Rickett}, B.~J. 1998, \apj, 507, 846

\bibitem[\protect\citeauthoryear{{Cordes} \& {Shannon}}{{Cordes} \&
  {Shannon}}{2010}]{2010arXiv1010.3785C}
{Cordes}, J.~M.,  \& {Shannon}, R.~M. 2010, arXiv:1010.3785

\bibitem[\protect\citeauthoryear{{Cordes} et~al.}{{Cordes}
  et~al.}{1990}]{1990ApJ...349..245C}
{Cordes}, J.~M., {Wolszczan}, A., {Dewey}, R.~J., {Blaskiewicz}, M.,  \&
  {Stinebring}, D.~R. 1990, \apj, 349, 245

\bibitem[\protect\citeauthoryear{{Craft}}{{Craft}}{1970}]{1970PhDT.........8C}
{Craft}, H.~D., Jr. 1970, Ph.D. thesis, CORNELL UNIVERSITY.

\bibitem[Damour 
\& Deruelle(1986)]{1986AIHS...44..263D} Damour, T., \& Deruelle, N.\ 1986, Ann.~Inst.~Henri Poincar{\'e} Phys.~Th{\'e}or., Vol.~44, No.~3, p.~263 - 292, 44, 263 

\bibitem[\protect\citeauthoryear{{Demorest}}{{Demorest}}{2007}]{2007PhDT........14D}
{Demorest}, P.~B. 2007, Ph.D. thesis, University of California, Berkeley

\bibitem[\protect\citeauthoryear{{Demorest}}{{Demorest}}{2011}]{2011MNRAS.416.2821D}
{Demorest}, P.~B. 2011, \mnras, 416, 2821

\bibitem[\protect\citeauthoryear{{Demorest} et~al.}{{Demorest}
  et~al.}{2013}]{2013ApJ...762...94D}
{Demorest}, P.~B., et~al. 2013, \apj, 762, 94

\bibitem[\protect\citeauthoryear{{Detweiler}}{{Detweiler}}{1979}]{1979ApJ...234.1100D}
{Detweiler}, S. 1979, \apj, 234, 1100

\bibitem[\protect\citeauthoryear{{DuPlain} et~al.}{{DuPlain}
  et~al.}{2008}]{2008SPIE.7019E..45D}
{DuPlain}, R., {Ransom}, S., {Demorest}, P., {Brandt}, P., {Ford}, J.,  \&
  {Shelton}, A.~L. 2008, in Society of Photo-Optical Instrumentation Engineers
  (SPIE) Conference Series, Vol. 7019, Society of Photo-Optical Instrumentation
  Engineers (SPIE) Conference Series

\bibitem[Espinoza et al.(2014)]{2014MNRAS.440.2755E} Espinoza, C.~M., 
Antonopoulou, D., Stappers, B.~W., Watts, A., 
\& Lyne, A.~G.\ 2014, \mnras, 440, 2755 

\bibitem[Espinoza et al.(2011)]{2011MNRAS.414.1679E} Espinoza, C.~M., Lyne, 
A.~G., Stappers, B.~W., \& Kramer, M.\ 2011, \mnras, 414, 1679 

\bibitem[\protect\citeauthoryear{{Finn} \& {Lommen}}{{Finn} \&
  {Lommen}}{2010}]{2010ApJ...718.1400F}
{Finn}, L.~S.,  \& {Lommen}, A.~N. 2010, \apj, 718, 1400

\bibitem[Foster 
\& Backer(1990)]{1990ApJ...361..300F} Foster, R.~S., \& Backer, D.~C.\ 1990, \apj, 361, 300 

\bibitem[Foster et al.(1993)]{1993ApJ...410L..91F} Foster, R.~S., 
Wolszczan, A., \& Camilo, F.\ 1993, \apjl, 410, L91 

\bibitem[\protect\citeauthoryear{{Grishchuk}}{{Grishchuk}}{2005}]{2005PhyU...48.1235G}
{Grishchuk}, L.~P. 2005, Physics Uspekhi, 48, 1235

\bibitem[\protect\citeauthoryear{{Hankins} \& {Rickett}}{{Hankins} \&
  {Rickett}}{1975}]{1975mcpr...14...55H}
{Hankins}, T.~H.,  \& {Rickett}, B.~J. 1975, in Methods in Computational
  Physics. Volume 14 - Radio astronomy, ed. B.~{Alder}, S.~{Fernbach}, \&
  M.~{Rotenberg}, Vol.~14, 55

\bibitem[\protect\citeauthoryear{{Hankins} \& {Rickett}}{{Hankins} \&
  {Rickett}}{1986}]{1986ApJ...311..684H}
{Hankins}, T.~H.,  \& {Rickett}, B.~J. 1986, \apj, 311, 684

\bibitem[\protect\citeauthoryear{{Hassall} et~al.}{{Hassall}
  et~al.}{2013}]{2013A&A...552A..61H}
{Hassall}, T.~E., et~al. 2013, \aap, 552, A61

\bibitem[\protect\citeauthoryear{{Hassall} et~al.}{{Hassall} et~al.}{2014}]{Has14}
{Hassall}, T.~E., et~al. 2014, {\it in preparation}

\bibitem[\protect\citeauthoryear{{Hellings} \& {Downs}}{{Hellings} \&
  {Downs}}{1983}]{1983ApJ...265L..39H}
{Hellings}, R.~W.,  \& {Downs}, G.~S. 1983, \apjl, 265, L39

\bibitem[\protect\citeauthoryear{{Hobbs}}{{Hobbs}}{2013}]{2013CQGra..30v4007H}
{Hobbs}, G. 2013, Classical and Quantum Gravity, 30, 224007

\bibitem[Hobbs et al.(2010)]{2010CQGra..27h4013H} Hobbs, G., Archibald, A., 
Arzoumanian, Z., et al.\ 2010, Classical and Quantum Gravity, 27, 084013 

\bibitem[\protect\citeauthoryear{{Hobbs}, {Edwards}, \& {Manchester}}{{Hobbs}
  et~al.}{2006}]{2006MNRAS.369..655H}
{Hobbs}, G.~B., {Edwards}, R.~T.,  \& {Manchester}, R.~N. 2006, \mnras, 369,
  655

\bibitem[\protect\citeauthoryear{{Hotan}, {van Straten}, \&
  {Manchester}}{{Hotan} et~al.}{2004}]{2004PASA...21..302H}
{Hotan}, A.~W., {van Straten}, W.,  \& {Manchester}, R.~N. 2004, \pasa, 21, 302

\bibitem[Jenet et al.(2004)]{2004ApJ...606..799J} Jenet, F.~A., Lommen, A., 
Larson, S.~L., \& Wen, L.\ 2004, \apj, 606, 799 

\bibitem[\protect\citeauthoryear{{Jones} et~al.}{{Jones}
  et~al.}{2015}]{Jon15}
{Jones}, G., et~al. 2015, {\it in
  preparation}

\bibitem[Karuppusamy et al.(2008)]{2008PASP..120..191K} Karuppusamy, R., 
Stappers, B., \& van Straten, W.\ 2008, \pasp, 120, 191 

\bibitem[\protect\citeauthoryear{{Karuppusamy} et~al.}{{Karuppusamy} et~al.}{2014}]{Kar14}
{Karuppusamy}, R., et~al. 2014, {\it in preparation}

\bibitem[\protect\citeauthoryear{{Kaspi}, {Taylor}, \& {Ryba}}{{Kaspi}
  et~al.}{1994}]{1994ApJ...428..713K}
{Kaspi}, V.~M., {Taylor}, J.~H.,  \& {Ryba}, M.~F. 1994, \apj, 428, 713

\bibitem[Keith et al.(2013)]{2013MNRAS.429.2161K} Keith, M.~J., Coles, W., 
Shannon, R.~M., et al.\ 2013, \mnras, 429, 2161 

\bibitem[\protect\citeauthoryear{{Kramer} \& {Champion}}{{Kramer} \&
  {Champion}}{2013}]{2013CQGra..30v4009K}
{Kramer}, M.,  \& {Champion}, D.~J. 2013, Classical and Quantum Gravity, 30,
  224009

\bibitem[\protect\citeauthoryear{{Kramer} et~al.}{{Kramer}
  et~al.}{1999}]{1999ApJ...526..957K}
{Kramer}, M., {Lange}, C., {Lorimer}, D.~R., {Backer}, D.~C., {Xilouris},
  K.~M., {Jessner}, A.,  \& {Wielebinski}, R. 1999, \apj, 526, 957

\bibitem[Kuzmin 
\& Losovsky(2001)]{2001A&A...368..230K} Kuzmin, A.~D., \& Losovsky, B.~Y.\ 2001, \aap, 368, 230 

\bibitem[\protect\citeauthoryear{{Lazio}}{{Lazio}}{2013}]{2013CQGra..30v4011L}
{Lazio}, T.~J.~W. 2013, Classical and Quantum Gravity, 30, 224011
  
\bibitem[Liu et al.(2014)]{Liu14} Liu, K., Desvignes, G., 
Cognard, I., et al.\ 2014, arXiv:1407.3827 

\bibitem[Lommen 
\& Backer(2001)]{2001ApJ...562..297L} Lommen, A.~N., \& Backer, D.~C.\ 2001, \apj, 562, 297 

\bibitem[\protect\citeauthoryear{{Lommen} \& {Demorest}}{{Lommen} \&
  {Demorest}}{2013}]{2013CQGra..30v4001L}
{Lommen}, A.~N.,  \& {Demorest}, P. 2013, Classical and Quantum Gravity, 30,
  224001

\bibitem[Lorimer et al.(2004)]{2004hpa..book.....L} Lorimer, D.~R., Kramer, 
M., Ellis, R., et al.\ 2004, Handbook of pulsar astronomy, by D.R.~Lorimer 
and M.~Kramer.~Cambridge observing handbooks for research astronomers, 
Vol.~4.~Cambridge, UK: Cambridge University Press, 2004

\bibitem[Madison et al.(2014)]{2014ApJ...788..141M} Madison, D.~R., Cordes, 
J.~M., \& Chatterjee, S.\ 2014, \apj, 788, 141 

\bibitem[Manchester 
\& IPTA(2013)]{2013CQGra..30v4010M} Manchester, R.~N., \& IPTA 2013, Classical and Quantum Gravity, 30, 224010 

\bibitem[\protect\citeauthoryear{{Manchester} et~al.}{{Manchester}
  et~al.}{2013}]{2013PASA...30...17M}
{Manchester}, R.~N., et~al. 2013, \pasa, 30, 17

\bibitem[\protect\citeauthoryear{{Matsakis}, {Taylor}, \& {Eubanks}}{{Matsakis}
  et~al.}{1997}]{1997A&A...326..924M}
{Matsakis}, D.~N., {Taylor}, J.~H.,  \& {Eubanks}, T.~M. 1997, \aap, 326, 924

\bibitem[\protect\citeauthoryear{{McLaughlin}}{{McLaughlin}}{2013}]{2013CQGra..30v4008M}
{McLaughlin}, M.~A. 2013, Classical and Quantum Gravity, 30, 224008

\bibitem[\protect\citeauthoryear{{Nita} et~al.}{{Nita}
  et~al.}{2007}]{2007PASP..119..805N}
{Nita}, G.~M., {Gary}, D.~E., {Liu}, Z., {Hurford}, G.~J.,  \& {White}, S.~M.
  2007, \pasp, 119, 805

\bibitem[Os{\l}owski et al.(2013)]{2013MNRAS.430..416O} Os{\l}owski, S., 
van Straten, W., Demorest, P., \& Bailes, M.\ 2013, \mnras, 430, 416 

\bibitem[\protect\citeauthoryear{{Os{\l}owski} et~al.}{{Os{\l}owski}
  et~al.}{2011}]{2011MNRAS.418.1258O}
{Os{\l}owski}, S., {van Straten}, W., {Hobbs}, G.~B., {Bailes}, M.,  \&
  {Demorest}, P. 2011, \mnras, 418, 1258

\bibitem[Pennucci et al.(2014)]{2014ApJ...790...93P} Pennucci, T.~T., 
Demorest, P.~B., \& Ransom, S.~M.\ 2014, \apj, 790, 93 

\bibitem[Rickett(1970)]{1970MNRAS.150...67R} Rickett, B.~J.\ 1970, \mnras, 
150, 67 

\bibitem[\protect\citeauthoryear{{Roy} et~al.}{{Roy}
  et~al.}{2010}]{2010ExA....28...25R}
{Roy}, J., {Gupta}, Y., {Pen}, U.-L., {Peterson}, J.~B., {Kudale}, S.,  \&
  {Kodilkar}, J. 2010, Experimental Astronomy, 28, 25

\bibitem[Sanidas et al.(2013)]{2013ApJ...764..108S} Sanidas, S.~A., Battye, 
R.~A., \& Stappers, B.~W.\ 2013, \apj, 764, 108 

\bibitem[\protect\citeauthoryear{{Sazhin}}{{Sazhin}}{1978}]{1978SvA....22...36S}
{Sazhin}, M.~V. 1978, \sovast, 22, 36

\bibitem[Sesana(2013)]{2013MNRAS.433L...1S} Sesana, A.\ 2013, \mnras, 433, 
L1 

\bibitem[\protect\citeauthoryear{{Sesana}, {Vecchio}, \& {Volonteri}}{{Sesana}
  et~al.}{2009}]{2009MNRAS.394.2255S}
{Sesana}, A., {Vecchio}, A.,  \& {Volonteri}, M. 2009, \mnras, 394, 2255

\bibitem[\protect\citeauthoryear{{Shannon} \& {Cordes}}{{Shannon} \&
  {Cordes}}{2010}]{2010ApJ...725.1607S}
{Shannon}, R.~M.,  \& {Cordes}, J.~M. 2010, \apj, 725, 1607

\bibitem[\protect\citeauthoryear{{Shannon} \& {Cordes}}{{Shannon} \&
  {Cordes}}{2012}]{2012ApJ...761...64S}
{Shannon}, R.~M.,  \& {Cordes}, J.~M. 2012, \apj, 761, 64

\bibitem[Shannon et al.(2014)]{2014MNRAS.443.1463S} Shannon, R.~M., 
Os{\l}owski, S., Dai, S., et al.\ 2014, \mnras, 443, 1463 

\bibitem[\protect\citeauthoryear{{Shannon} et~al.}{{Shannon}
  et~al.}{2013}]{2013Sci...342..334S}
{Shannon}, R.~M., et~al. 2013, Science, 342, 334

\bibitem[\protect\citeauthoryear{{Stairs} et~al.}{{Stairs}
  et~al.}{2002}]{2002ApJ...581..501S}
{Stairs}, I.~H., {Thorsett}, S.~E., {Taylor}, J.~H.,  \& {Wolszczan}, A. 2002,
  \apj, 581, 501
  
\bibitem[Stappers et 
al.(2011)]{2011A&A...530A..80S} Stappers, B.~W., Hessels, J.~W.~T., Alexov, A., et al.\ 2011, \aap, 530, A80 

\bibitem[\protect\citeauthoryear{{Starobinski{\v i}}}{{Starobinski{\v
  i}}}{1979}]{1979JETPL..30..682S}
{Starobinski{\v i}}, A.~A. 1979, Soviet Journal of Experimental and Theoretical
  Physics Letters, 30, 682

\bibitem[\protect\citeauthoryear{{Stinebring}}{{Stinebring}}{2013}]{2013CQGra..30v4006S}
{Stinebring}, D. 2013, Classical and Quantum Gravity, 30, 224006

\bibitem[\protect\citeauthoryear{{Taylor}}{{Taylor}}{1992}]{1992RSPTA.341..117T}
{Taylor}, J.~H. 1992, Royal Society of London Philosophical Transactions Series
  A, 341, 117

\bibitem[\protect\citeauthoryear{{van Haarlem} et~al.}{{van Haarlem}
  et~al.}{2013}]{2013A&A...556A...2V}
{van Haarlem}, M.~P., et~al. 2013, \aap, 556, A2

\bibitem[\protect\citeauthoryear{{van Haasteren} \& {Levin}}{2010}]{2010MNRAS.401.2372V} van Haasteren, R., \& Levin, Y.\ 2010, \mnras, 401, 2372 

\bibitem[\protect\citeauthoryear{{van Haasteren} et~al.}{{van Haasteren} et~al.}{2011}]{2011MNRAS.414.3117V}
{van Haasteren}, R., et~al. 2011, \mnras, 414, 3117

\bibitem[\protect\citeauthoryear{{van Straten}}{{van
  Straten}}{2004}]{2004ApJS..152..129V}
{van Straten}, W. 2004, \apjs, 152, 129

\bibitem[van Straten et 
al.(2012)]{2012AR&T....9..237V} van Straten, W., Demorest, P., \& Oslowski, S.\ 2012, Astronomical Research and Technology, 9, 237 

\bibitem[Verbiest et al.(2009)]{2009MNRAS.400..951V} Verbiest, J.~P.~W., 
Bailes, M., Coles, W.~A., et al.\ 2009, \mnras, 400, 951

\bibitem[\protect\citeauthoryear{{Walker}, {Demorest}, \& {van Straten}}{{Walker} et~al.}{2013}]{2013ApJ...779...99W}
{Walker}, M.~A., {Demorest}, P.~B.,  \& {van Straten}, W. 2013, \apj, 779, 99

\bibitem[You et al.(2007)]{2007ApJ...671..907Y} You, X.~P., Hobbs, G.~B., 
Coles, W.~A., Manchester, R.~N., \& Han, J.~L.\ 2007, \apj, 671, 907 

\bibitem[\protect\citeauthoryear{{Zhu} et~al.}{{Zhu} et~al.}{2015}]{Zhu15}
{Zhu}, W., et~al. 2015, {\it in preparation}

\end{thebibliography}

\end{document}